\documentclass[11pt]{article}
\usepackage{graphicx,color} 
\usepackage{amsmath,mathrsfs,amssymb}
\usepackage{natbib}
\setcitestyle{authoryear,open={(},close={)}}

\usepackage[english]{babel}
\usepackage[T1]{fontenc}

\newcommand{\DEF}{\overset{\mathrm{def}}{=}}

\newcommand{\dmat}{\mathrm{d}}

\renewcommand{\theenumi}{(\kern -0.15ex{\roman{enumi}})}

\renewcommand{\th}{${}^\mathrm{th}$ }

\begin{document}
\onecolumn

\noindent \textbf{\LARGE Drowning by numbers: topology and physics in fluid dynamics} \\[1cm]
Amaury Mouchet \\
Laboratoire de Math\'ematiques et de Physique Th\'eorique, Universit\'e  Fran\c{c}ois Rabelais de Tours, CNRS (UMR 7350),
 F\'ed\'eration Denis Poisson, 37200 Tours, France\\ [2cm]

Since its very beginnings, topology has forged strong links with
physics and the last Nobel prize in physics, awarded in 2016 to
Thouless, Haldane and Kosterlitz ``for theoretical discoveries of
topological phase transitions and topological phases of matter'',
confirmed that these connections have been maintained up to
contemporary physics.  To give some (very) selected illustrations of
what is, and still will be, a cross fertilization between topology and
physics\footnote{A more general review is proposed by \citet{Nash99a}
  and a systematic presentation on the topological concepts used by
  physicists can be found in \citep{Nakahara90a}.  }, hydrodynamics
provides a natural domain through the common theme offered by the
notion of vortex, relevant both in classical (\S\;2) and in quantum
fluids (\S\;3). Before getting into the details, I will sketch in
\S\;1 a general perspective from which this intertwining between
topology and physics can be appreciated: the old dichotomy between
discreteness and continuity, first dealing with antithetic thesis,
eventually appears to be made of two complementary sides of a single
coin.

\section{The arena of the discrete/continuous dialectic}

One century after Thales of Miletus had proposed that water was the natural principle of
all things, the first atomists Leucippus and Democritus advocated for a discrete
conception of matter. The existence of an ultimate lower limit of divisibility, materialised by the atoms,
may have been  a logical answer to the Zeno's paradoxes~(\citealt[chap.~VIII]{Stokes71a}; \citealt[chap.~I]{Bell06a}).  
In some westernmost banks of the Mediterranean sea, the Pythagorean school was concerned by
a line of thought following quite an opposite direction: the discovery of the irrational numbers
counterbalanced the  conception of a universe exclusively
driven by the integer and rational---in the original acception of the word---numbers. 
For twenty-five centuries, the dialectic between continuity and  discreteness has never stopped nurturing
natural philosophy. 

At our daily life scales, the ones for which the brains have been
shaped by Darwinian evolution\footnote{In modern times physics and
  chemistry were not, by far, the only scientific disciplines to be
  shaken by violent debates between discrete and continuous schools;
  in the \textsc{xix}\th century Lyell's uniformitarianism in
  geology, by contrast with catastrophism, had an important influence
  on the young Darwin. By the way, one can notice that the binary
  opposition between discreteness and continuity provides by itself a
  meta self-referring epistemological dichotomy, so to speak.},
discreteness appears to be an inevitable way for intelligence to model
the world\footnote{However, neurology shows that numerical cognition is more
  analogical than numerical: beyond few units, the numbers are encoded
  and treated by the brain as fuzzy entities~\cite[specially part~I
    and chap.~9]{Dehaene97b}. }. Furthermore, operationally speaking,
any measurement is reduced, in the last resort, to a reproducible 
counting~\cite[\S\;1.1]{Thouless98a}.  
Etymologically, ``discrete'',
``critical'', ``criterion'', and ``discernment'' share the same greek
root~$\kappa\rho\acute\iota\nu\omega$
(\textit{kr}$\acute{\text{\it\={\i}}}$\textit{n\=o}, to
judge)\footnote{The etymology lines of these words can be easily traced back with
  \texttt{www.wiktionary.org}.}. However, the boundaries of
macroscopic objects, considered both in space and time, remain
inevitably blurred. For instance, consider one cherry; through
absorption and desorption, a perpetual exchange of matter takes place
at small scales on the skin of the cherry, and no one can really
identify with a precision of one second the time when this cherry has
appeared from a blossom or destroyed by natural
deterioration\footnote{In a contribution to the previous volume of
  this series~\cite[\S\;5]{Mouchet15a}\nocite{Emmer+15a} I have tried
  to show how symmetries play a crucial role in the process of
  abstraction and conceptualisation of a macroscopic object like a
  cherry.}. This ambiguity was known from antiquity and supply
the sorites paradox (what is the minimum number of
grains in a heap of sand?)---and the paradox of the ship of Theseus
(Plutarch asks if, after decades of restauration, once her last plank
has been replaced, the ship remains the same Theseus's ship~\cite[The
  life of Theseus \S\;XXIII.1]{Plutarch14a}).

In the second part of the \textsc{xix}\th century, experiments allowed
to move the debate beyond speculations into the microscopic world. In
the same movement, mathematics saw the emergence of a new discipline,
topology, where were identified some \emph{discrete}
classifications---first in geometry, then in analysis and algebra---up
to \emph{continuous} invertible transformations (homeomorphisms).  The
integer numbers upon which the classes of, say, graphs, knots,
surfaces, fixed points of a flow, critical points of a real map, are
discriminated provide, by essence, a robust quantization; they are
topological invariant.  To put it in a nutshell, there cannot be
``half a hole''. The dimension of a space\footnote{In fractal
  geometry, the Hausdorff dimension of a set, which can be
  irrationnal, is not preserved by a homeomorphism.  }, its
connectedness ($\pi_0$), its homotopy groups ($\pi_1$, $\pi_2$ and
more generally~$\pi_n$), the signature of the Hessian of a function at
a critical point, are examples of such discrete quantities.
 
In the beginning of the \textsc{xx}\th century, quantum physics refuted
so masterfully the Leibniz continuity principle (\textit{Nature does
  not make jumps}) that it bears this claim in its very name. The
general rule---known by Pythagoreans for music---according to which a
stable wave in a bounded domain has its frequencies quantized (that
is, function of integer numbers) now applied at a fundamental level to
the Schr\"odinger waves, which described the states of elementary
particles, when bounded.  The discrete classification of chemical
elements successfully proposed in 1869 by Mendeleev and the discrete
spectral lines corresponding to the Balmer series, the Paschen series,
the Lyman series etc. observed in radiation, could be explained within a unifying scheme
offered by quantum theory.  Eventhough it appears that each atomic
energy level has actually a continuous bandwidth, due to the coupling
to the electromagnetic field whose scattering states belong to a
continuum (the photon has no mass), it is nevertheless quantum theory
that confered to ``being an integer'' a genuine physical property. So
far, neither the quantification of the spin nor the quantification of
the electric charge, say, can be seen as an approximation of a continuous
model and the analogous of the Mendeleiev table in the Standard Model
contains a finite number of species of elementary particles---about
twenty, non counting as distinct a particle from its associated
antiparticle---characterised by a handful of quantum
numbers\footnote{The discrete character of some observable properties
  is all the more strengthened that there exists some superselection
  rules that make irrelevant any continuous superposition of states
  differing by some discrete values of this observable. }. Many
attempts have been made for finding a topological origin of these
quantum numbers, one of the motivation being that topological invariance
is much harder to break than symmetry invariance. In condensed matter, topology offers a protection against
the effects of impurities or out-of-control perturbations and therefore participates 
to the reproductibility and the fiability of measurements~\cite[\S\;1.3]{Thouless98a}.
The seminal attempt in this direction is Dirac's model of magnetic
monopole~\citep{Dirac31a} whose existence would imply the quantization
of the electric charge; however, so far, all the quantizations that
have been explained find their
root in \emph{algebraic} properties of the symmetry groups used to build a
basis of quantum states\footnote{Topological properties of these Lie
  groups, obviously their dimensions but also their compactness, their connectedness and their simple
  connectivity, do play a role but the algebraic commutation relations of
  their generators remain the main  characteristics, which are local ones, that
  allow to build the irreducible representations defining the
  one-particle states. } (in the absence of evidence of elementary
magnetic monopoles, the fact that the electric charges appear to be
always an integer multiple of one unit remains mysterious).

Despite these (temporary?) failures of finding topological
rather than algebraic roots for the discrete characteristics of what
appears to be elementary particles, the quantum theory of fields
offers the possibility of describing some collective effects of those
particles whose stability is guaranteed by topological
considerations. There exists some configurations of a macroscopic
number of degrees of freedom that cannot be created or destroyed by a
smooth transformation without passing through an intermediate state
having a macroscopic, and therefore redhibitory, energy. Depending on the dimension of the space
and of the field describing the model, several such \emph{topological
  defects} can be considered (point, lines or surfaces) and have been
observed in various condensed states~\citep[chap.~9]{Chaikin/Lubensky95a}
including, of course, the quantum fluids where the defects are
characterised by quantized numbers that can be interpreted as
topological invariants.  Vortices, which will be the object of the next two sections,
 provide typical examples of such
topological defects along a line in a 3-dimensional space or localised
at one point in a 2-dimensional space for a complex scalar field (or a
real bidimensional vector field). 
Under certain circumstances, these
collective effects share many properties with the so-called ordinary
particles. Since, theoretically, the distinction between the quasi-particles and
particles appears, after all, to be just a matter of convention on the
choice of the vacuum and of the particles that are considered to be
elementary, one may have the secret hope that at a more fundamental
level, having the Standard Model as an effective theory, topology
shall have the next, but presumably not the last, word.

\section{Classical vortices}

\begin{flushright}
\textit{\dots when I first opened my eyes upon the wonders of the whirlpool\dots}\\
Edgar Allan Poe. \emph{A Descent into the Maelstr\"om} (1841).
\end{flushright}

\subsection{How vortices participate to the dynamics of the world according to Leonardo and Descartes}

\begin{figure}[!ht]
\begin{center}
\includegraphics[width=\textwidth]{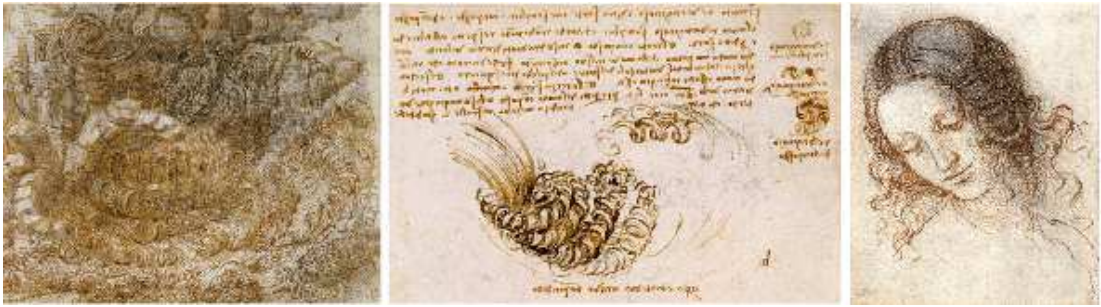}
\caption{\label{fig:leonardtourbillons} Left) folio~w12380, A D\'eluge,
  $\sim$1517-18.  Center) folio~w12663r, Studies of flowing water,
  $\sim$1510-13. Right) folio~w12518, Head of Leda, $\sim$1504-1506.
  \texttt{Wikimedia Foundation}. On the folio~w12579r Leonardo has
  drawn four studies of vortex alleys formed in water behind a
  parallelepipedic obstacle and writes \textit{Nota il moto del
    livello dell'acqua, il quale fa a uso de' capelli, che hanno due
    moti, de' quali l'uno attende al peso del vello, l'altro al
    liniamento delle volte: cos\`{\i} l'acqua ha le sue volte
    revertiginose, delle quali una parte attende al impeto del corso
    principale, l'altra attende al moto incidente e refresso.}
  (Observe the movement of the surface of
  water, like hair which has two movements, one due to its weight, the
  other following the lines of the curls: thus water has whirling
  eddies, in part following the impetus of the main stream, in part
  following the incidental and reversed motion, folio w12579r, trad.~\textsc{am}). }
\end{center}
\end{figure}

By strong contrast with the still, rather mineral, backgrounds of his paintings,
Leonardo da Vinci's interest for the dynamics of water is manifest
in his drawings and writtings all along his life.
Vortices in water, in air, and even in blood \cite[\S\;3.3]{Pasipoularides10a}, were a recurrent source of 
fascination for him\footnote{\citet{Gombrich69a}\nocite{OMalley69a} saw in the exuberance of the terms
used by Leonardo and in the profusion of his drawings an attempt to classify
the vortices, a line of investigations he kept in mind throughout his life.}.
Not only as esthetical motifs (fig.~\ref{fig:leonardtourbillons}), not only because of their crucial role
for understanding hydraulics and fly, not only because they inspired him fear as a disordered manifestation of
flooding or deluge, but also because they provided a central key for
 his global conception of the dynamics of the world:
\textit{l'acqua, vitale omore della terreste macchina, mediante il suo natural calore
si move.} (water, vital humour of the terrestrial machine, moves by means of its natural heat)\footnote{Folio H95r, whose facsimile and transcription can be found on \texttt{www.leonardodigitale.com}.}
\cite[Chap.~\textit{Une science en mouvement}]{Arasse97a}.

More than a century later, most probably without any influence from Leonardo, Descartes
put the vortices in the very core of his cosmological model. Rejecting 
the atomist concept of a vacuum separating matter \cite[part~II, 16th principle]{Descartes1644a},
he writes
\begin{quotation}
\textit{[\dots] putandum est, non tantum Solis \& Fixarum, sed totius etiam coeli
materiam fluidam esse.}\\
([\dots] we think that not only the 
 matter of the Sun and of the Fixed Stars is fluid but also is
the matter of all the sky, trad. \textsc{am})
\flushright \cite[\S~III.24 p. 79]{Descartes1644a}
\end{quotation}
Being aware of the proper rotation of the Sun (it takes 26 days for
the sunspots to complete one turn \cite[\S~III.32 p.~83]{Descartes1644a})
and of the different orbital period of the planets, he pursues further
the hydrodynamical analogy
\begin{quotation}
\textit{[\dots] putemus totam materiam coeli in qua Planetae
  versantur, in modum cuiusdam vorticis, in cuius centro est Sol,
  assidue gyrare, ac eius partes Soli viciniores celerius moveri quam
  remotiores [\dots]}\\ ([\dots] we think that all the matter of
the sky, in which the Planets turn, rotates like a vortex with the Sun
at its center; that the parts near the Sun move faster than the remote
ones [\dots],
trad. \textsc{am})
\flushright \cite[\S~III.30 pp. 81-82]{Descartes1644a}
  \end{quotation}
\begin{figure}[!ht]
\begin{center}
\includegraphics[width=7cm]{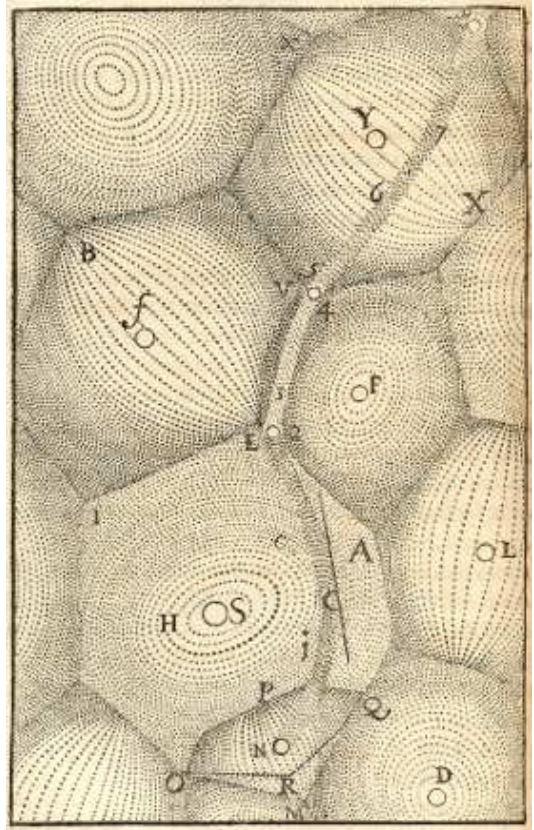}
\caption{\label{fig:descartestourbillons} Descartes'vortex-based cosmology. Each star denotes by F, D, etc.
is at the center of a vortex. The Sun is denoted by S \protect\cite[\S~III.23 p.~78]{Descartes1644a}. }
\end{center}
\end{figure}
Descartes'model was overuled by Newton's theory planetary motion but,
somehow, in contemporary astrophysics, vortices are still present---in
a complete different way, of course, from Descartes'--- and triggered
by gravitational field acting through the interstellar vacuum: one
may think of protoplanetary accretion disks (turbulence plays a crucial
role, in particular in the initial molecular cloud for explaining the
scattered births of stars) and, at much larger scales, of galaxies,
cosmic whirlpools spinning around a giant black hole.

\subsection{Accompanying the birth of topology in the \textsc{xix}\th century}

\begin{figure}[!ht]
\begin{center}
\parbox[c]{10cm}{\includegraphics[width=10cm]{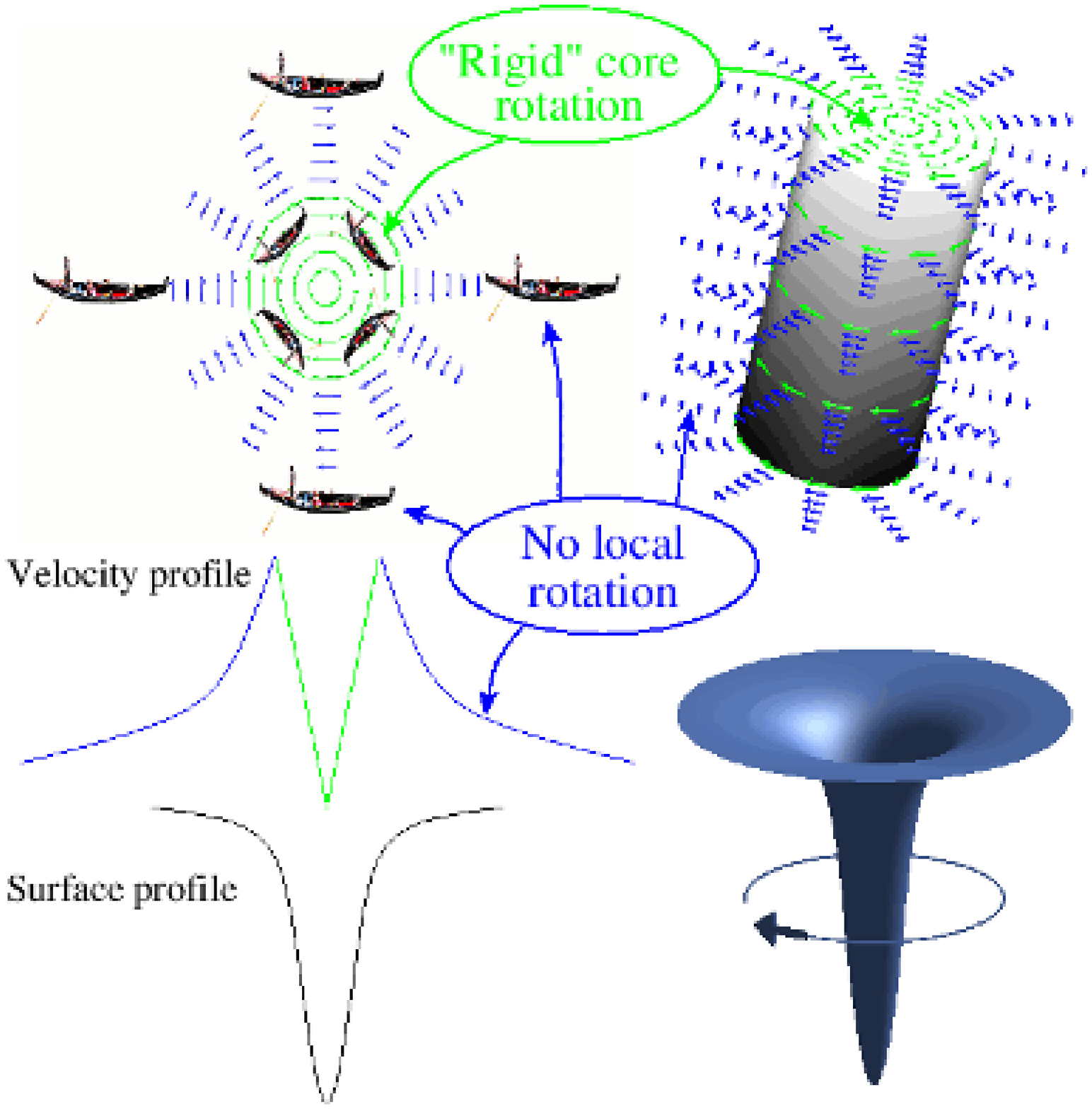}}
\parbox[c]{4cm}{\includegraphics[width=4cm]{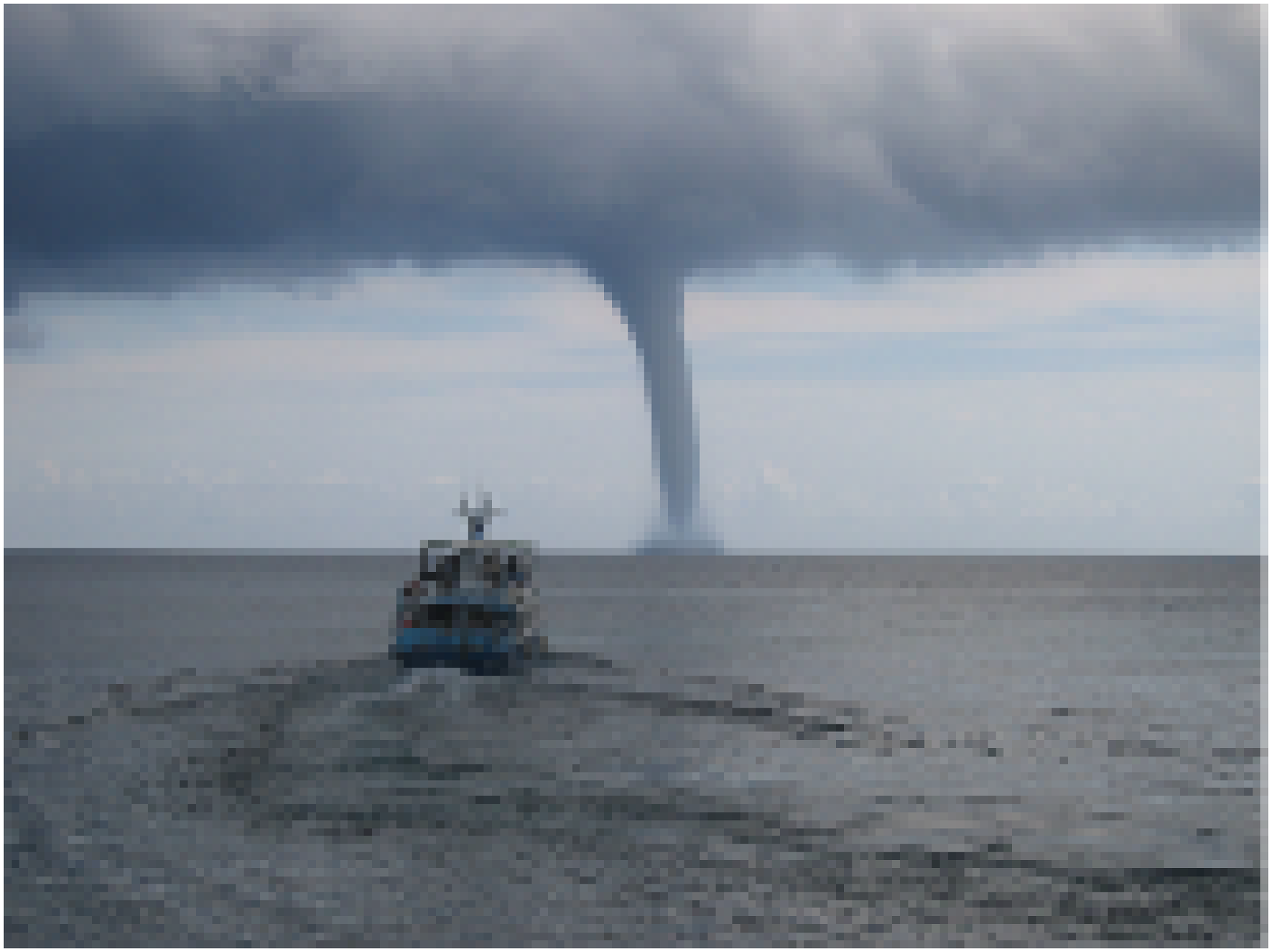}
\includegraphics[width=4cm]{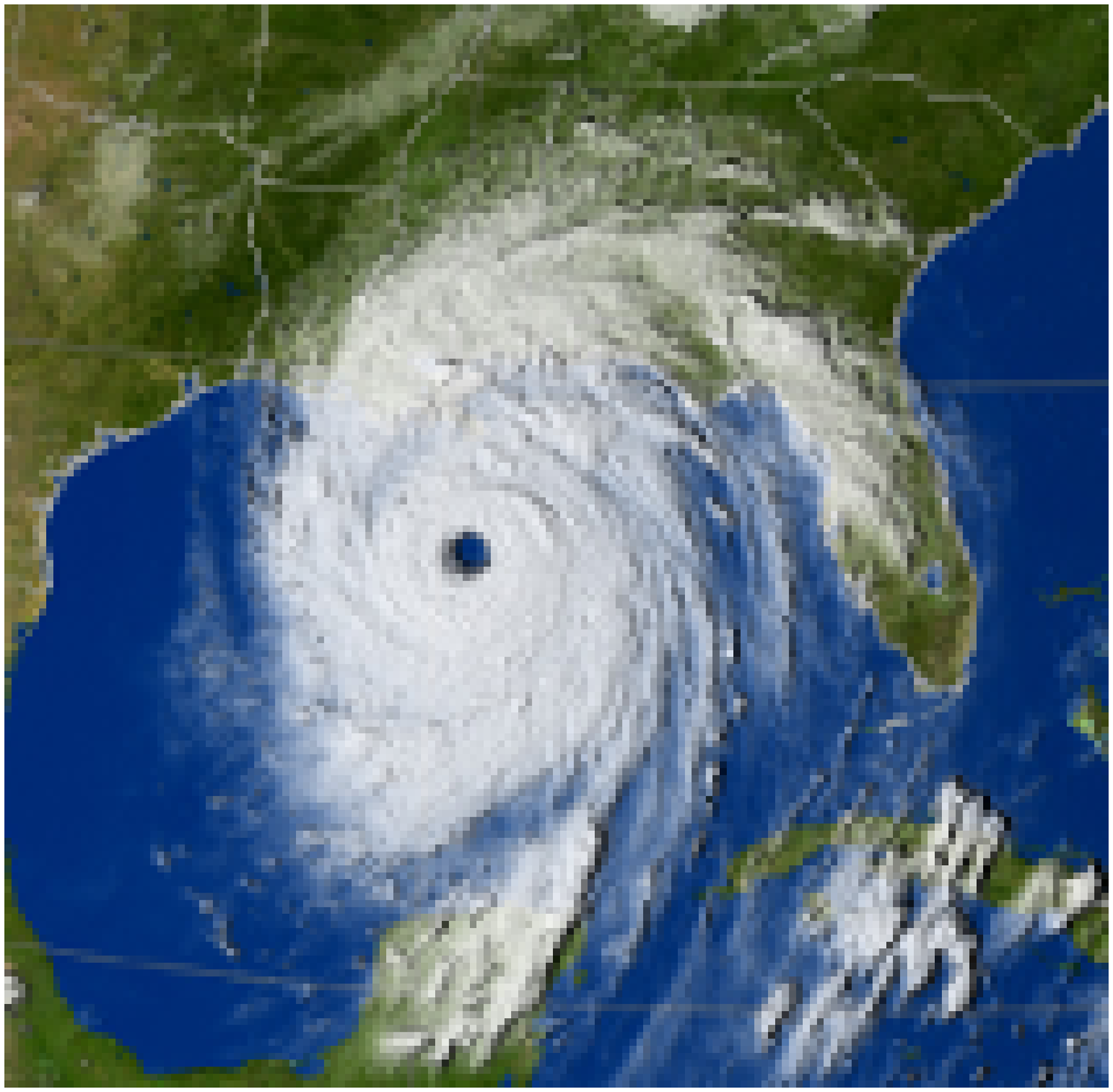}
\includegraphics[width=4cm]{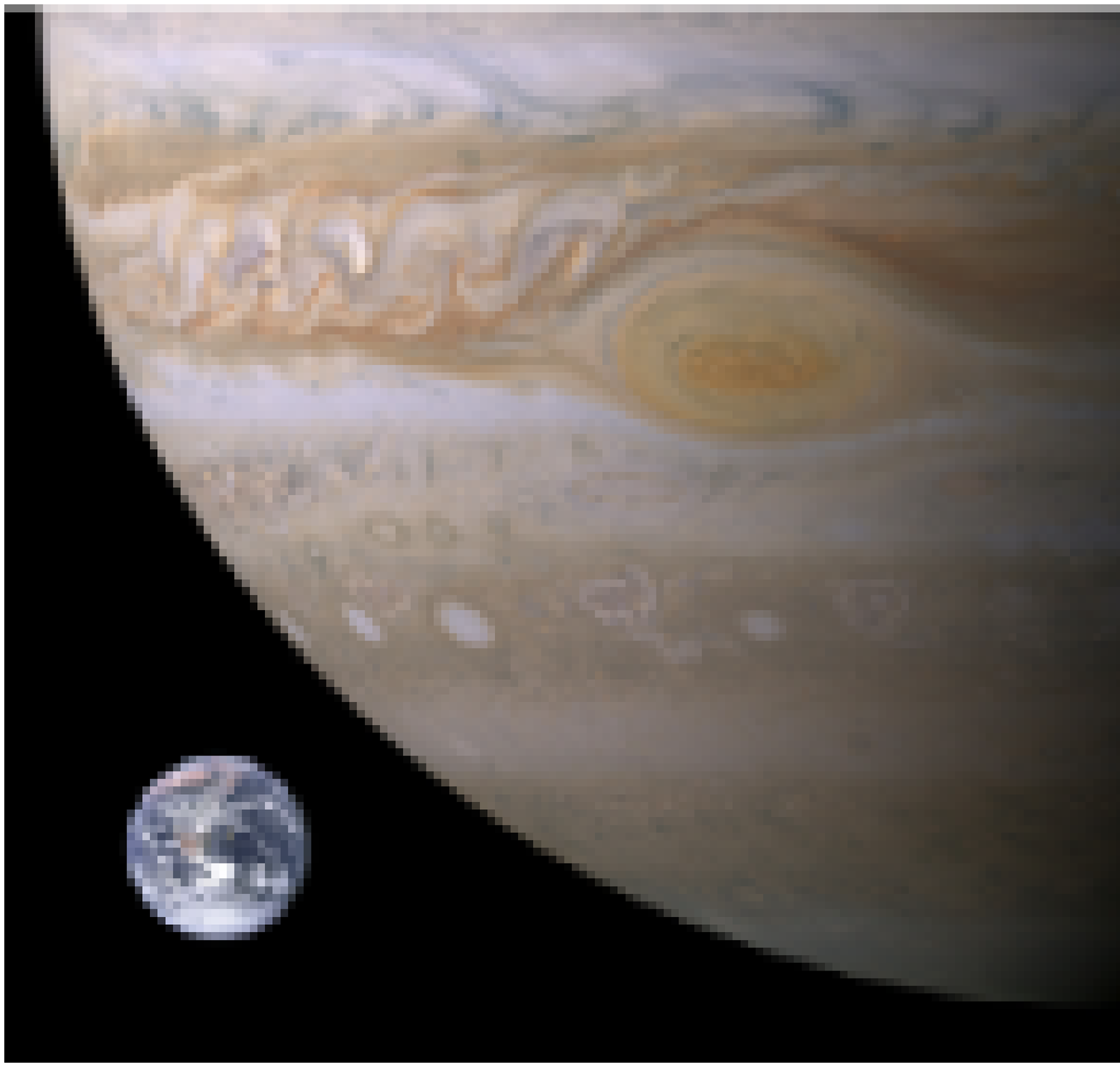}
\includegraphics[width=4cm]{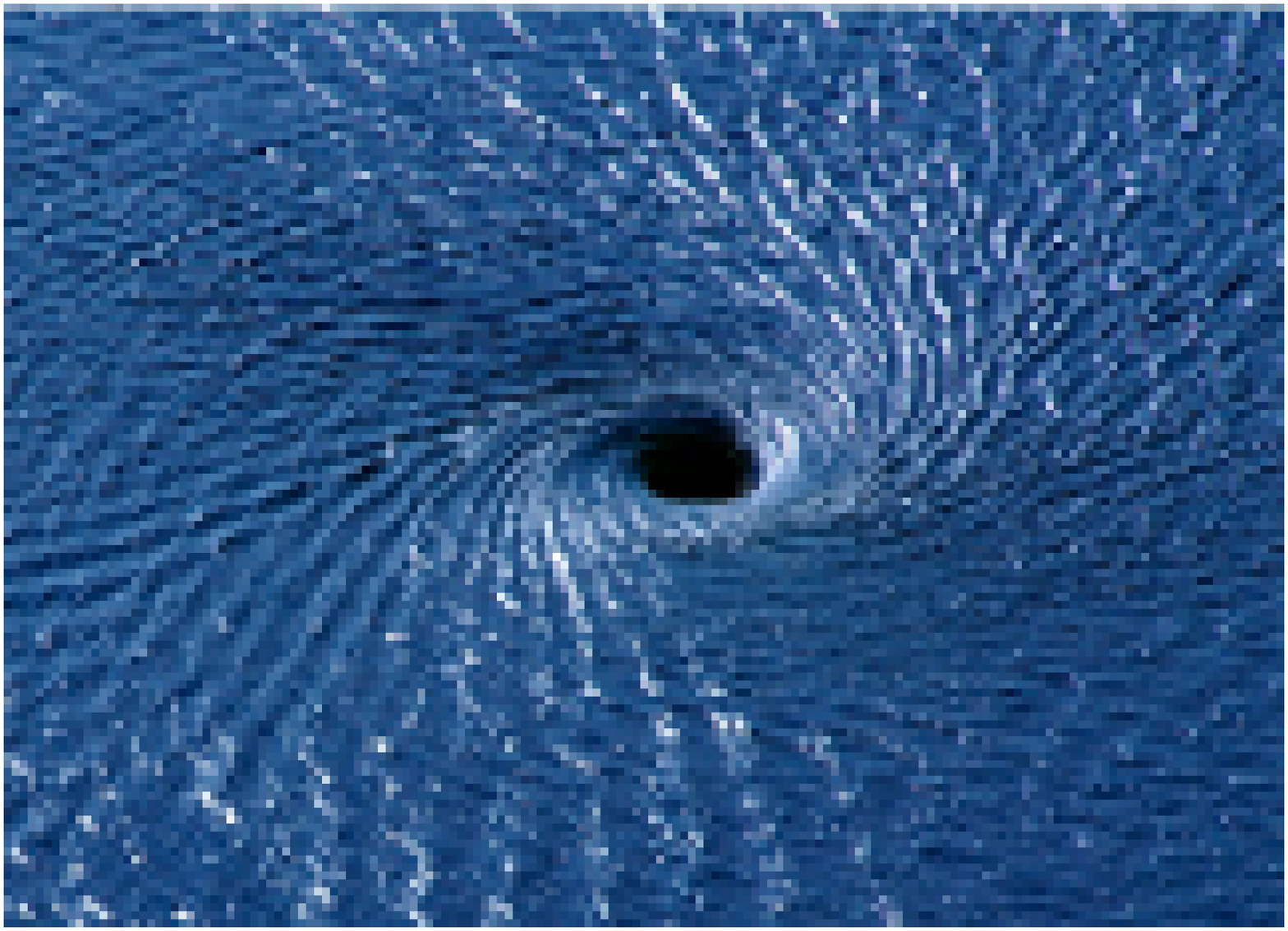}}
\caption{\label{fig:vortexstructure} The same year Helmholtz published
  his seminal memoir, the simplest model of vortex was explicitely
  proposed by Rankine in \protect\cite[\S\S\;629-633]{Rankine1858a}
  who refers to some previous theoretical analysis made by the
  engineer and physicist James Thomson, inventor of the vortex wheel
  and brother of William. The vorticity~\eqref{eq:omega} is constant
  and uniform inside a cylinder---in green, where the fluid rotates as a solid core
  and the particles
  rotate around themselves (the axis of the gondola rotates)---and zero outside---in blue, where the fluid
  particles do not rotate around themselves (the axis of the gondola keeps the same direction). When coming closer to
  the axis of the vortex, the velocity increases with the inverse of
  the distance outside the cylindrical core (and then producing a
  spiral-like shape) and then linearly gets to zero inside the
  core. In a more or less realistic way, Rankine's vortex models
  hurricanes, tornados or simply water going down a plughole (image
  credit: wikipedia, \textsc{noaa}).  }
\end{center}
\end{figure}

\begin{figure}[!ht]
\begin{center}
\parbox[c]{12cm}{\includegraphics[width=5cm]{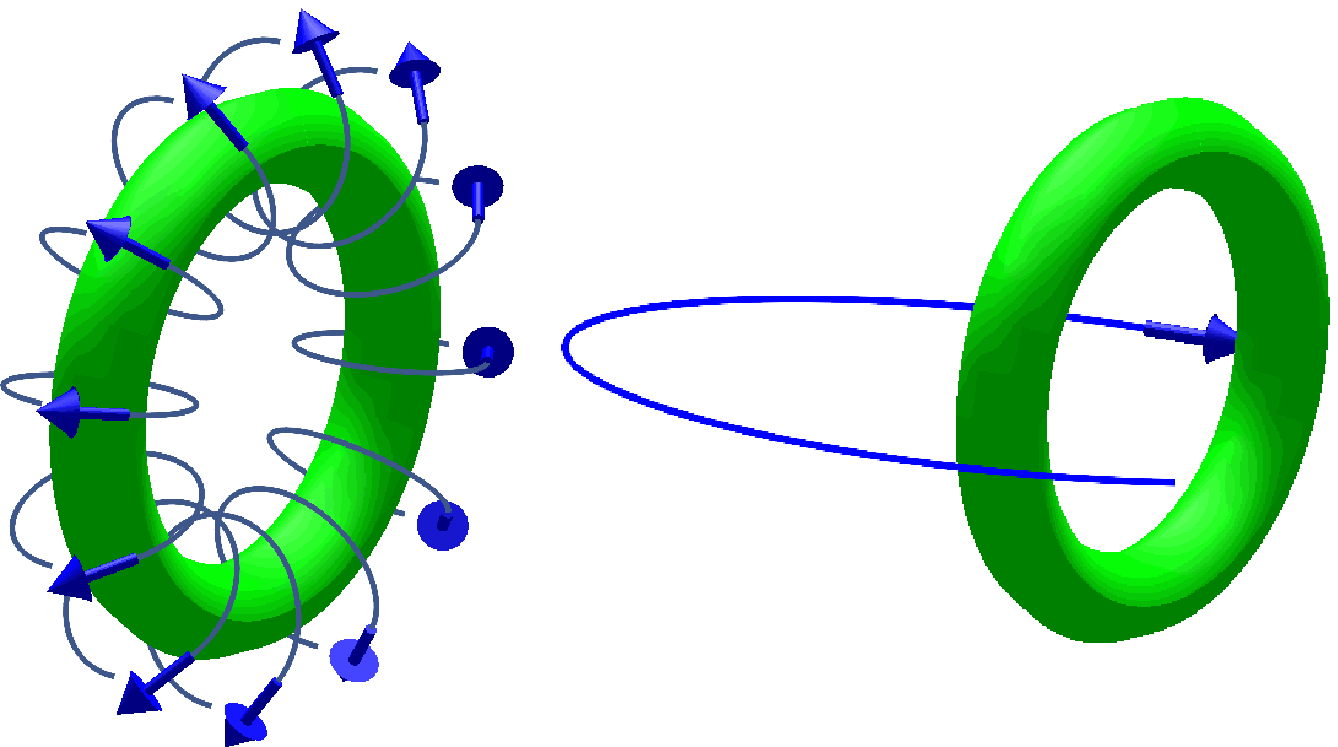}
\includegraphics[width=7cm]{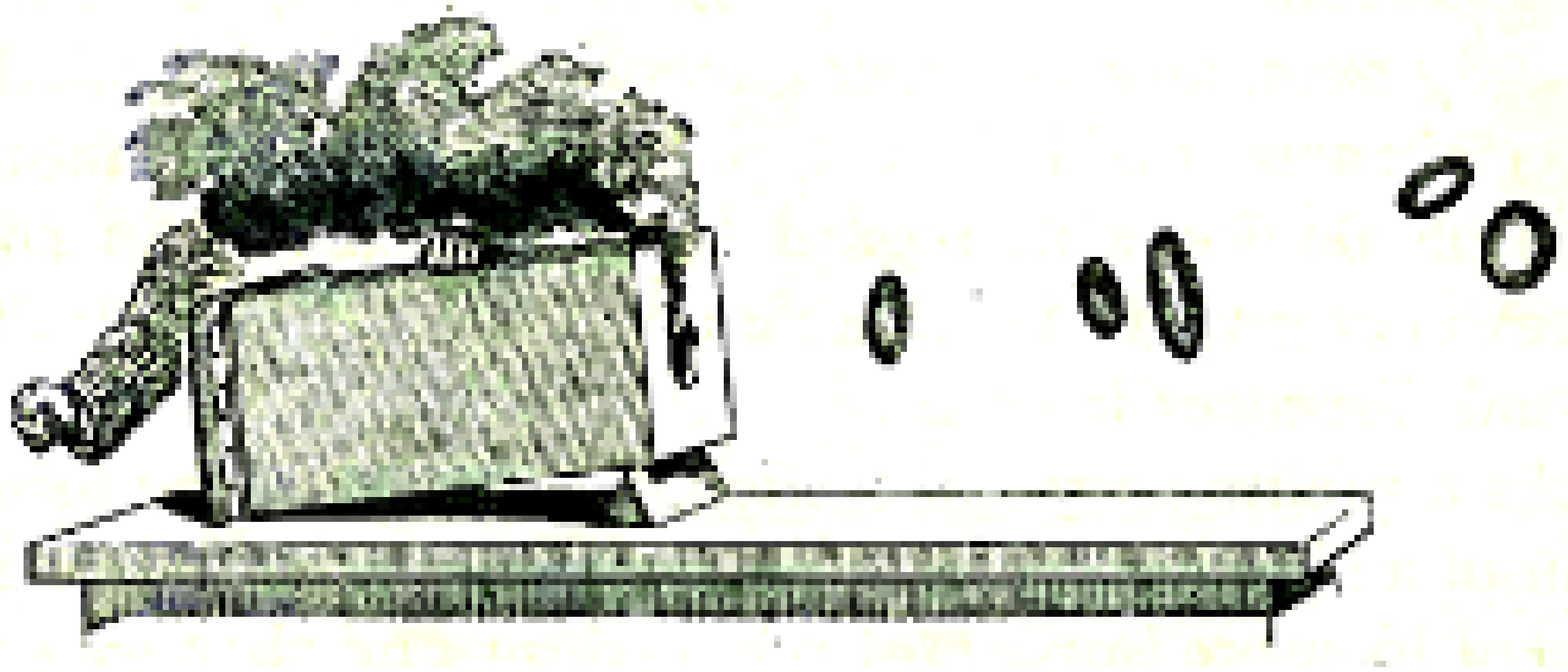}}
\includegraphics[width=4cm]{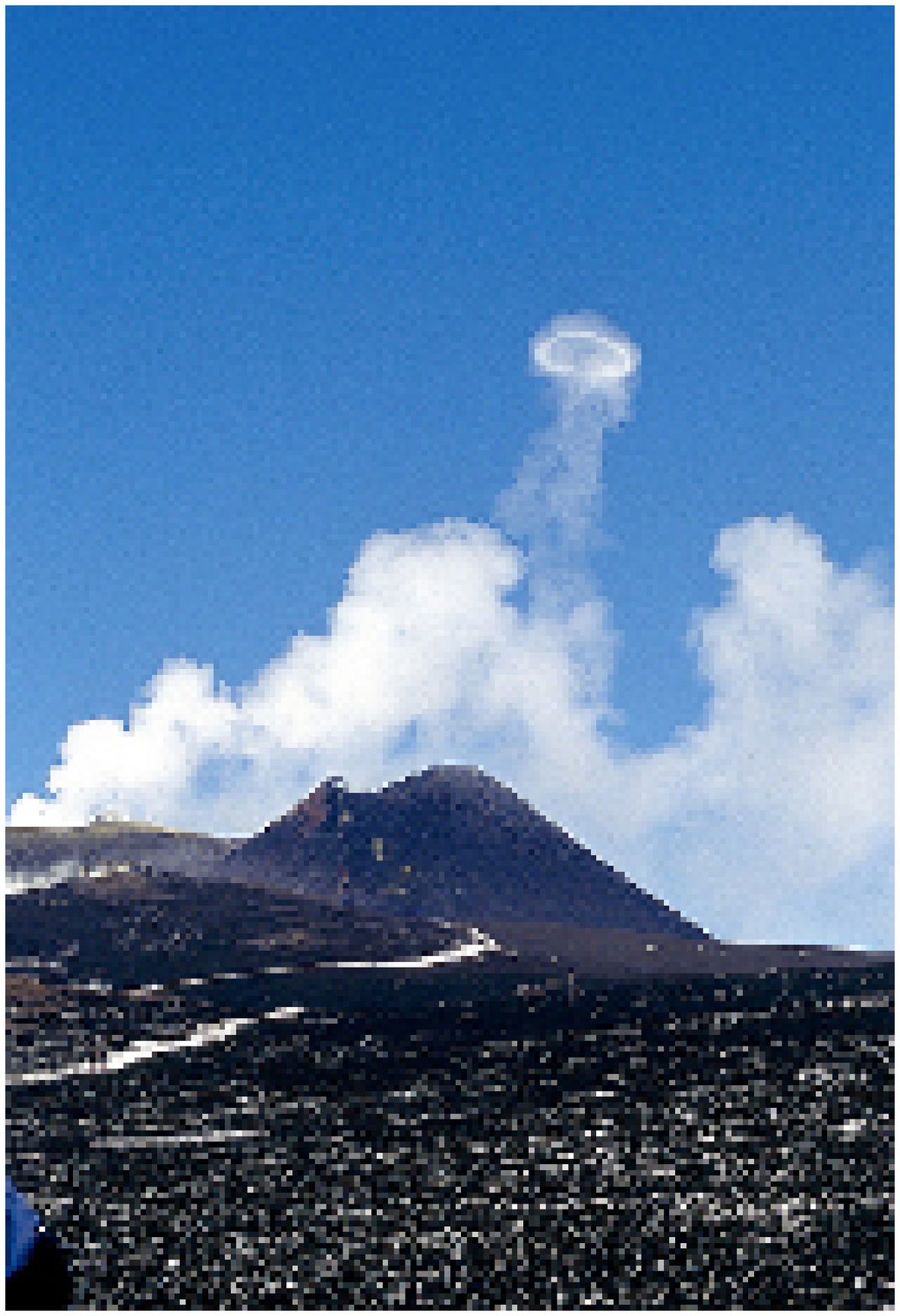}\ \ \ \ 
\includegraphics[width=4.15cm]{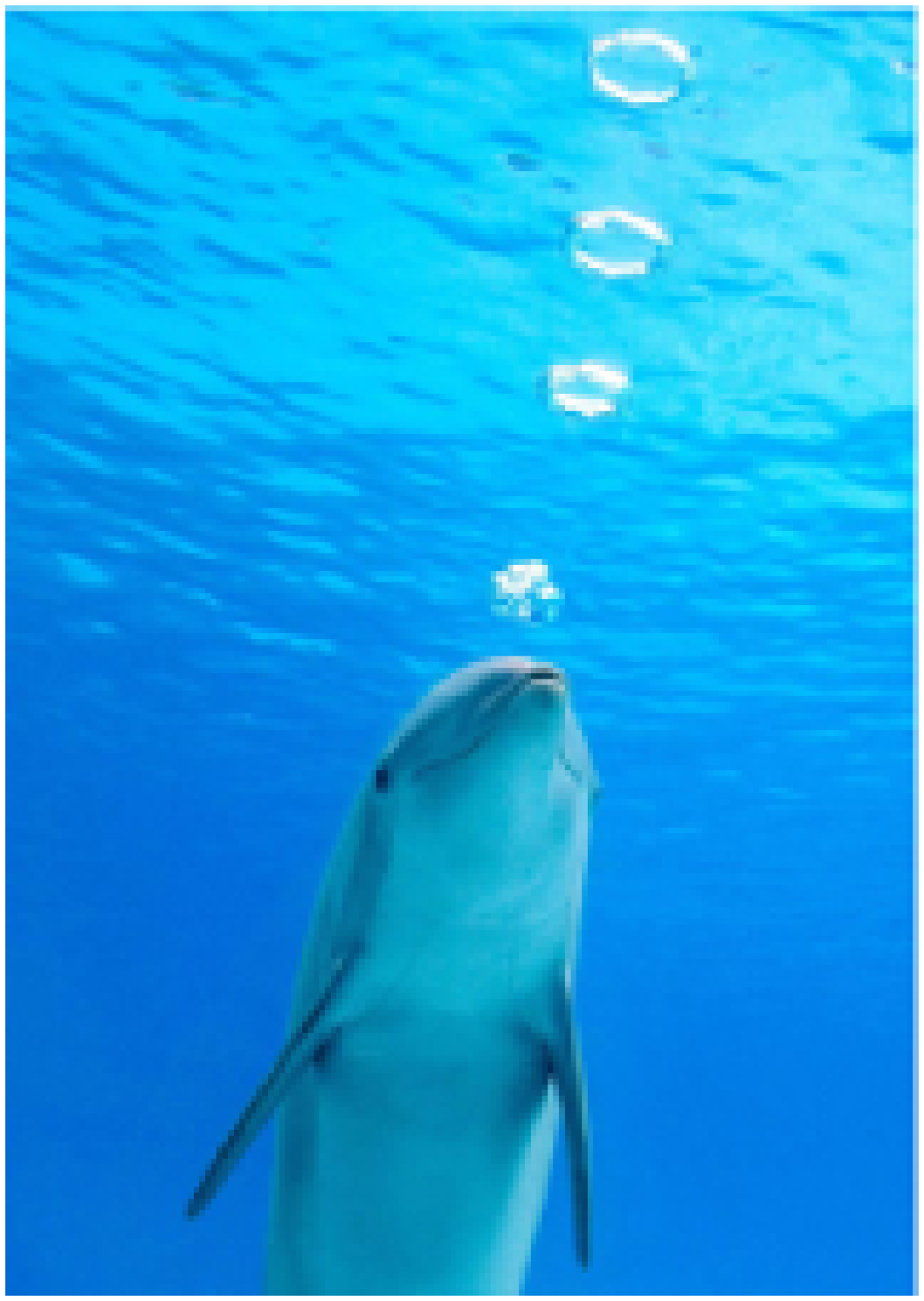}
\caption{\label{fig:vortexanneau} When some vortex lines are bended into a circular tube (in green), each portion of the ring
is dragged in the same direction by the fluid whose motion is induced by the other parts of the ring.
As a result, a global translation perpendicular to the ring occurs. Helmholtz'study of the dynamics of the rings and the tango played
by two interacting rings moving in the same direction, see footnote~\ref{fn:ringtango} p.~\pageref{fn:ringtango}, 
can be visualised with Tait's smoke box (upper right taken from \protect\cite[p.~292]{Tait1884a}). In exceptional circumstances vapour rings
can be naturally produced by vulcanos (the lower left photograph is taken at Etna by the vulcanologist Boris Behncke, \textsc{invg}-Osservatorio Etneo). Dolfins and whales are able to produce vortex rings in water (lower right from youtube).  } 
\end{center}
\end{figure}

His study of the physical properties of organ pipes led Helmholtz to
scrutinize the motion of the air near sharp obstacles and the
influence of viscosity.  The memoir he published in German in 1858 on
the subject had a decisive influence on the physicists of the Scottish
school including Maxwell, Rankine, Tait and Thomson (who was ennobled
in 1892 as Lord Kelvin), all the more that Tait translated it into
English in 1867 under the title \textit{On the integrals of the
  hydrodynamical equations, which express
  vortex-motion}~\citep{Helmholtz1867a}.  Inspired by the parallel
between mechanics of continuous media and
electromagnetism \citep[chap.~4]{Darrigol05a}, Helmholtz showed that,
given a field of velocities~$\vec{v}$, its curl, the vorticity field,
\begin{equation}\label{eq:omega}
  \vec{\omega}=\overrightarrow{\mathrm{curl}}\,\vec{v}
\end{equation} 
is a vector field proportional to the local rotation vector of the
fluid. Helmholtz introduced the notion of vortex line (a curve tangent
to~$\vec{\omega}$ at each of its points) and vortex filament/tube (a
bunch of vortex lines) and proved that during its evolution each
vortex line follows the motion of the fluid. The dynamical equation
of~$\vec{\omega}$ allowed him to study precisely the dynamics of
straight (fig.~\ref{fig:vortexstructure}) and circular vortex tubes (fig.~\ref{fig:vortexanneau}). A thin vortex ring whose
radius~$R$ is much larger than the radius of the cross section of the
tube that defines it moves perpendicularly to its plane with the
velocity of its center increasing with~$R$\footnote{\label{fn:ringtango}In particular,
  when two rings moving along the same direction get close,
 the flow created around the leading ring tends to shrink
  the following one which, conversely, generates a flow that tends to expend
 the ring ahead. Therefore the leading ring slows down while the second
one is sped up until it
  overtakes the former by passing through it, and the
  role of the rings are exchanged. This tango, predicted and observed
  by Helmholtz, is described in the end of his 1858 memoir.}.  Based
on the similar mathematical problem arose in electrostatics and
magnetostatics, Helmholtz understood that the topology of the
irrotational part of the flow was essential to determine
\emph{globally} the velocity potential~$\alpha$: in the set of the
points~$P$ where~$\vec{\omega}(P)=0$ one can always \emph{locally}
define a scalar field~$\alpha$ such
that
\begin{equation}\label{eq:gradphi}
  \vec{v}=\overrightarrow{\mathrm{grad}}\,\alpha
\end{equation}
but
\begin{quotation} If we consider [a vortex-filament] as always reentrant either within or without
the fluid, the space for which [equation~\eqref{eq:gradphi}] holds is
complexly connected, since it remains single if we conceive surfaces
of separation through it, each of which is completely bounded by a
vortex-filament. In such complexly connected spaces a
function~[$\alpha$] which satisfies the above equation can have more
than one value ; and it must be so if it represents currents
reentering, since the velocity of the fluid outside the
vortex-filaments are proportional to the differential coefficients
of~[$\alpha$], and therefore the motion of the fluid must correspond
to ever increasing values of~[$\alpha$]. If the current returns to
itself, we come again to a point where it formely was, and find there
a second greater value of~[$\alpha$]. Since this may occur
indefinitely, there must be for every point of such a
complexly-connected space an infinite number of distinct values
of~[$\alpha$] differing by equal quantities like those
of~$\tan^{-1}\frac{x}{y}$, which is such a many-valued function
[\dots].

\hfill \cite[\S\;3, translation by Tait]{Helmholtz1867a}.
\end{quotation}   

\begin{figure}[!ht]
\begin{center}
\includegraphics[width=10cm]{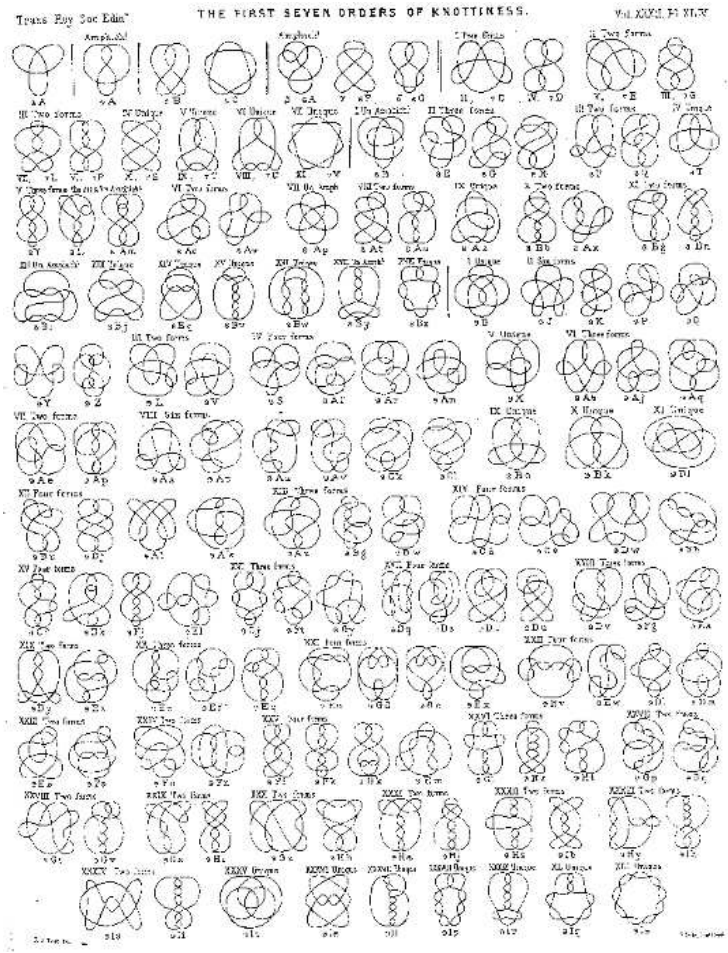}
\caption{\label{fig:Taitknots} List of knots up to the seventh order established by \citet[Plate XLIV between p.~338 \& 339]{Tait1884a}.}
\end{center}
\end{figure}

The topological properties of vortices can also be understood from
what is now known as Kelvin's circulation theorem
\cite[\S\;59d]{Thomson1869a} which unified Helmholtz results: in an inviscid (no viscosity),
barotropic (its density is a function of pressure only) fluid, the
flux of the vorticity
\begin{equation}\label{eq:circulationv}
  \Gamma=\int_{\mathscr{S}}\vec{\omega}\cdot\dmat\vec{S}=\int_{\partial\mathscr{S}}\vec{v}\cdot\dmat\vec{l}
\end{equation}
through a surface~$\mathscr{S}$ following the motion of the fluid---or
equivalently, according to Stokes' theorem, the circulation of the
velocity through the boundary~$\partial\mathscr{S}$
of~$\mathscr{S}$---is constant. As a consequence, we recover Helmholtz
statement that the non simple connectedness of the space filled by the
irrotational part of the flow, i.e. the complementary of the vortex
tubes, prevents the existence of a continuous globally-defined~$\alpha$
and the circulation~$\Gamma$ depends on the homotopy class of the
loop~$\mathscr{C}=\partial\mathscr{S}$. In such an ideal fluid, the
vortex lines were therefore topologically stable and Thomson's saw in
this stability a key for the description of atomic properties without
referring to the corpuscular image inheritated from the atomists of
antiquity, which was a too suspicious philosophy for Victorian
times~\cite[\S\S\;2 and~9]{Kragh02a}\footnote{Some smoothness into the
  atom had already been introduced by Rankine in 1851 with his
  hypothesis of \emph{molecular vortices} according to which ``each
  atom of matter consists of a nucleus or central point enveloped by
  an elastic atmosphere, which is retained in its position by
  attractive forces, and that the elasticity due to heat arises from
  the centrifugal force of those atmospheres, revolving or oscillating
  about their nuclei or central points''
  \cite[\S\;2]{Rankine1851a}. It is worth noting that Rankine
  acknowledges the pertinence of William Thomson's comments on the first
  version of this 1851's proposal.  }. Since vortex tubes cannot cross
transversaly\footnote{But, it seems that neither Helmholtz nor Thomson
  have considered the possibility of a longitudinal merging of vortex
  tubes, forming a trousers-like shape \cite[in particular
    fig.~6]{VelascoFuentes07a}.} otherwise it is easy to find
a~$\mathscr{C}$ that does not satisfy Kelvin's theorem, the knot
formed by a closed vortex tube and the intertwinning between several
such closed loop remain topologically invariant.
\begin{quotation} 
The absolute permanence of the rotation, and the unchangeable relation you have
proved between it and the portion of the fluid once acquiring such motion in a
perfect fluid, shows that if there is a perfect fluid all through space, constituting the
substance of all matter, a vortex-ring would be as permanent as the solid hard atoms
assumed by Lucretius and his followers (and predecessors) to account for the permanent
properties of bodies (as gold, lead, etc.) and the differences of their characters.
Thus, if two vortex-rings were once created in a perfect fluid, passing through one
another like links of a chain, they never could come into collision, or break one
another, they would form an indestructible atom; every variety of combinations
might exist.
\flushright Thomson to Helmholtz, January 22, 1867, quoted by \cite[p.~38]{Kragh02a}.
\end{quotation}
The theory of the vortex atoms offered to Thomson the possibility of making concrete his
 long-standing intuition of a continuous conception of the world, as he
had confessed it to Stokes 
\begin{quotation} 
Now I think hydrodynamics is to be the root of all physical science,
and is at present second to none in the beauty of mathematics.
\flushright Thomson to Stokes, December, 20, 1857,\\ quoted in
\cite[p.~35]{Kragh02a}
\end{quotation}
Despite the physical failure of Thomson's ambitious aim
\citep{Silliman63a,Epple98a,Kragh02a}\footnote{As far as classical
  hydrodynamics is concerned, some progress have been made in the
  \textsc{xx}\th century with, for instance, the identification of new
  integrals of motion constructed from topological invariants like the
  Calugareanu helicity \citep{Moffatt08a}\nocite{Borisov+08a} ; experimentally some not
  trivial knotted vortices could be produced only recently
  \citep{Kleckner/Irvine13a}.}, the identification of topological
invariants on knots, upon which the classification of atoms and
molecules would have been based, and the classification of the knots
by Tait (see Fig.~\ref{fig:Taitknots} for instance) remains a groundbreaking mathematical work, with direct
repercussions in contemporary topology.

One of the Thomson's greatest hopes, while spectroscopy was gathering
more and more precise data, was to explain the origin of the discrete
spectral lines with `` [\dots] one or more fundamental periods of
vibration, as has a stringed instrument of one or more strings
[\dots]''~\cite[p.~96]{Thomson1867a}. One cannot prevent to find an
echo of this motivation in modern string theory where ``each particle
is identified as a particular vibrational mode of an elementary
microscopic string''~\cite[\S\;1.2]{Zwiebach04a}---see also \cite[in
  particular \S\;19]{Cappelli+12a}.  Not without malice, \citet{Kragh02a}
was perfectly right to qualify Thomson's dream as a ``Victorian theory
of everything''.

\section{Quantum vortices}

\subsection{Topological origin of quantized flux in quantum fluids}

Unlike what occurs in classical fluids where viscosity eventually make
the vortices smoothly vanish, quantum fluids provide a state of
matter, much more similar to ideal fluids, where vortices are
strongly protected from dissipative processes.  Indeed, at low
temperature, particles can condensate into a collective quantum state
where transport can be dissipationless: this is one of the main
characteristics of superconductivity (discovered in solid mercury
below~4K by Onnes in 1911), superfluidity (discovered in liquid
Helium-4 below~2K by Kapitsa and Allen \& Misener in 1938), and
Bose-Einstein condensate of atoms (discovered for rubidium below
170~nK by Cornell \& Wieman and Ketterle in 1995)\footnote{One can
find many
textbooks at different levels and more or less specialised
to one type of quantum fluids. To get an introductory bird's-eye view
on quantum fluids and other matters in relation to statistical physics, my personal
taste go to \citep{Chaikin/Lubensky95a}, \citep{Huang87a} and the particularly
sound, concise, and pedagogical \citep{Sator/Pavloff16a} (in French).
}. 
There is a second
reason, of topological origin,  that reinforces the stability of the vortices in quantum fluids:
the scalar field~$\alpha$ whose gradient is proportional to the current 
 is not a simple mathematical intermediate
as in the classical case (see~\eqref{eq:gradphi}) but acquires the more physical status of
being a phase (an angle) that may be measured in interference experiments like
in the Aharonov-Bohm effect. As a consequence, on any closed loop~$\mathscr{C}$, the circulation~$\Gamma$ given by~\eqref{eq:circulationv} 
has to be an integer multiple of~$2\pi$:
\begin{equation}\label{eq:winding}
  w[\mathscr{C}]\DEF\frac{1}{2\pi}\int_{\mathscr{C}}\overrightarrow{\mathrm{grad}}\,\alpha\cdot\dmat\vec{l}\in\mathbb{Z}\;.
\end{equation}
Since smooth transformations cannot provoque discrete jumps, $w$~is
therefore topologically protected.  In other words, the flux
of~$\overrightarrow{\mathrm{curl}}\,\vec{v}$---which keeps its
physical interpretation of being a vorticity in superfluids as well as
in Bose-Einstein condensates of atoms, whereas it represents a
magnetic field in superconductors\footnote{Compare~\eqref{eq:omega}
  with the relation~$\vec{B}=\overrightarrow{\mathrm{curl}}\,\vec{A}$
  between the (gauge) vector potential~$\vec{A}$ and the magnetic
  field~$\vec{B}$.}---is quantized and naturally leads to elementary
vortices carrying a unit flux quantum.  As a matter of fact, the
quantum fluid state is described by a complex field~$\psi=|\psi|{\large
  \mathrm{e}}^{\mathrm{i}\alpha}$ (the order parameter)
and~$w[\mathscr{C}]\neq0$ denotes a singularity of the order parameter
on any surface~$\mathscr{S}$ whose boundary
is~$\mathscr{C}$. Vortices constitute a particular case of what is
generally called a \emph{topological defect} whose dimension depends
on the dimension of the order parameter and on the dimension of the
space.
\begin{figure}[!ht]
\begin{center}
\parbox[c]{5cm}{\includegraphics[width=5cm]{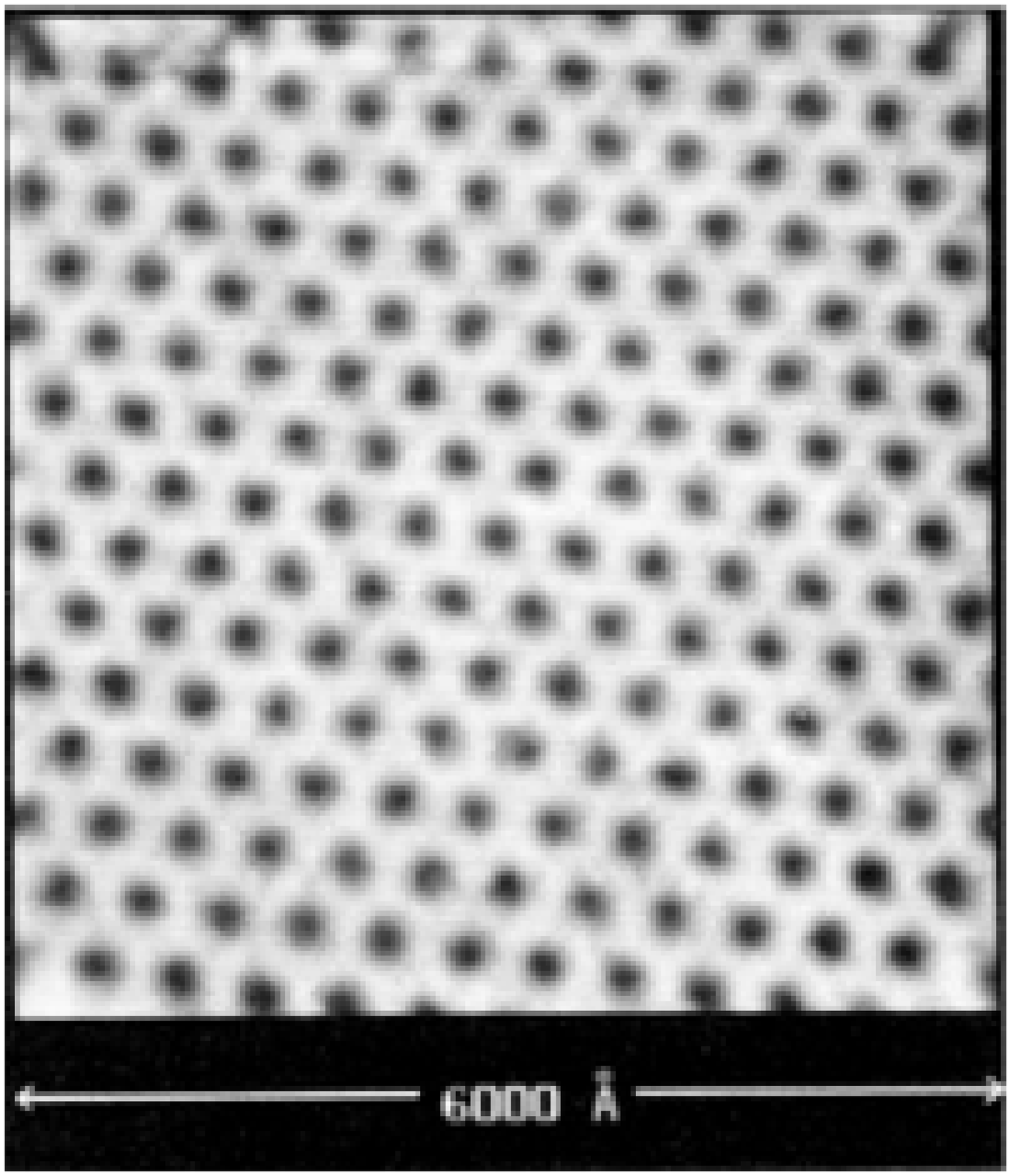}}
\parbox[c]{7cm}{\includegraphics[width=7cm]{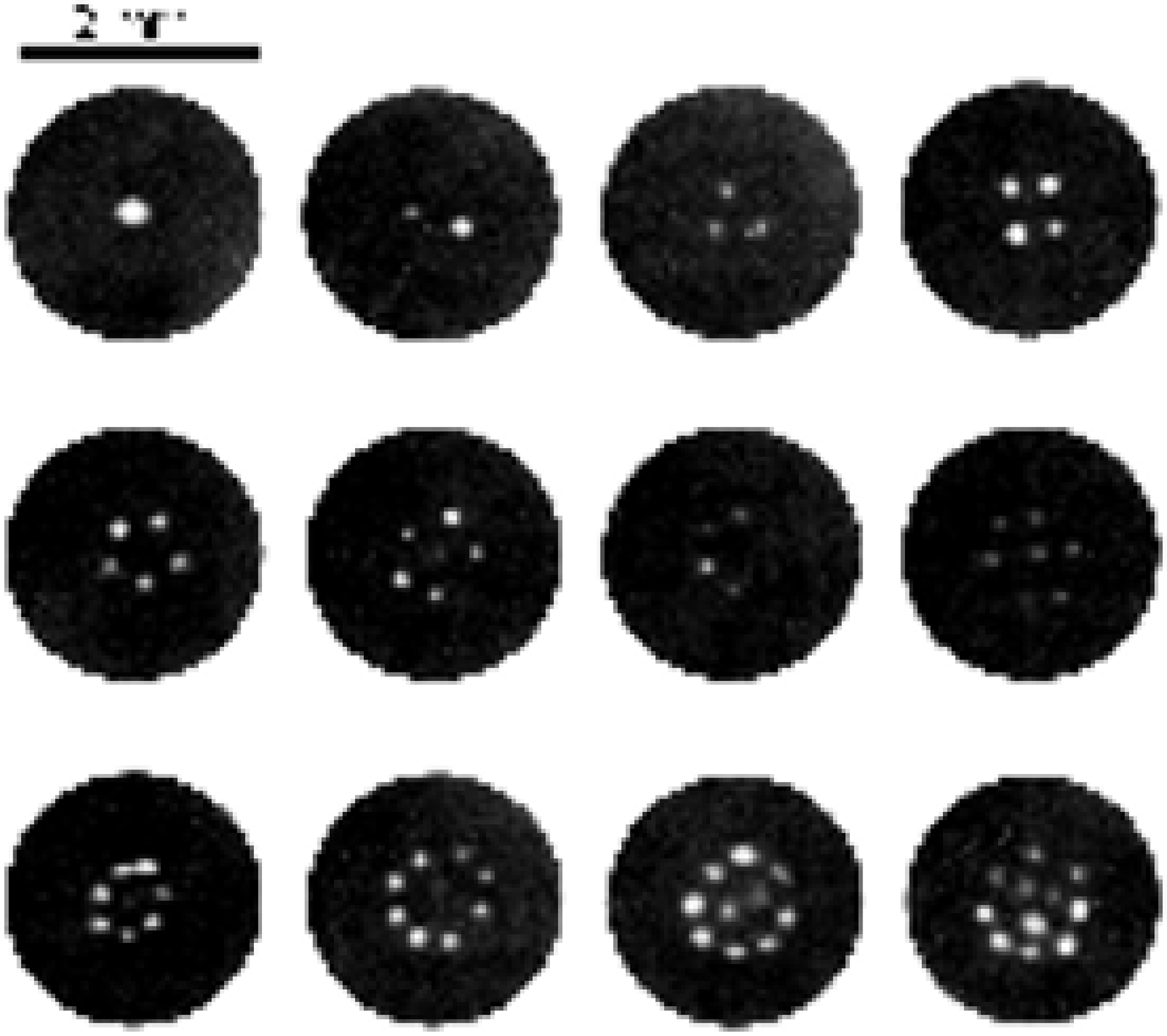}}
\parbox[c]{12cm}{\includegraphics[width=12cm]{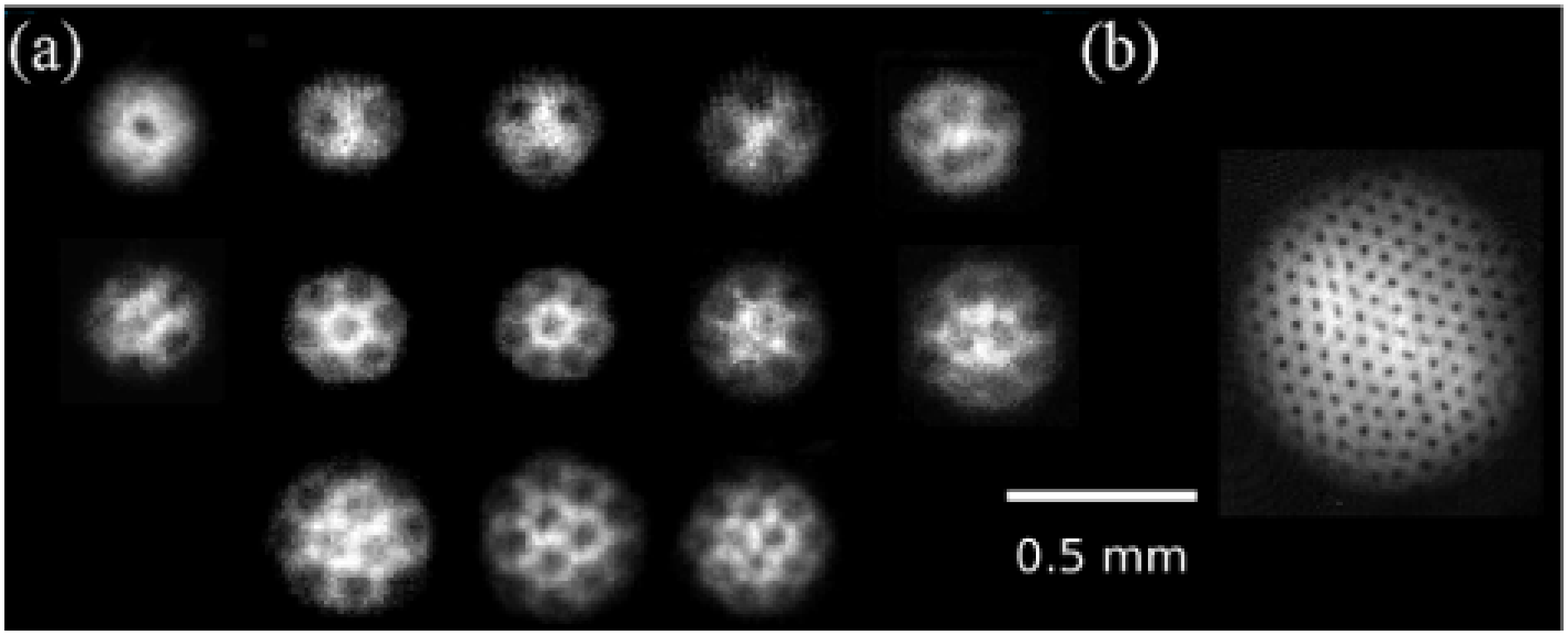}}
\caption{\label{fig:abrikosov} Up left)~Abrikosov lattice of vortices
  in a superconductor \protect\cite[fig.~2]{Hess+89a}.  Up right)
  Vortices in superfluid helium \cite[fig.~2]{Yarmchuk+79a}.  Below)
  Vortices in a rotating Bose-Einstein condensate obtained by
  a)~Dalibard's group (\protect\citealt[fig.~1]{Madison+00a} \&
  \protect\citealt[fig.~4]{Chevy/Dalibard06a}); b)~Ketterle's group
  \protect\cite[fig.~4c]{Raman+01a}. \copyright European Physical
  Society and American Physical Society. }
\end{center}
\end{figure}
  At microscopic scales, very much like in the Rankine model,
the vortex is made of a core outside
which~$\overrightarrow{\mathrm{curl}}\,\vec{v}=0$; the
vorticity/magnetic lines are trapped inside the core where the density
of the superfluid~$|\psi|^2$ tends to zero at its center. Not only,
these vortices have been observed in all the three types of
superfluids mentioned above but also the triangular lattice they form
to minimize the (free) energy due to an effective repulsion between
them first predicted by \citet{Abrikosov57a}, see fig~\ref{fig:abrikosov}).
When the fluctuations of~$|\psi|$ in space and time are negligible,
notably at sufficiently low temperatures, the quantum fluid is
essentially described by the phase~${\mathrm{e}}^{\mathrm{i}\alpha}$
or equivalently by a bidimensional vector of unit norm oriented at
angle~$\alpha$ with respect to a given direction
(fig.~\ref{fig:modeleXY_def}).
\begin{figure}[!bh]
\begin{center}
\includegraphics[width=8cm]{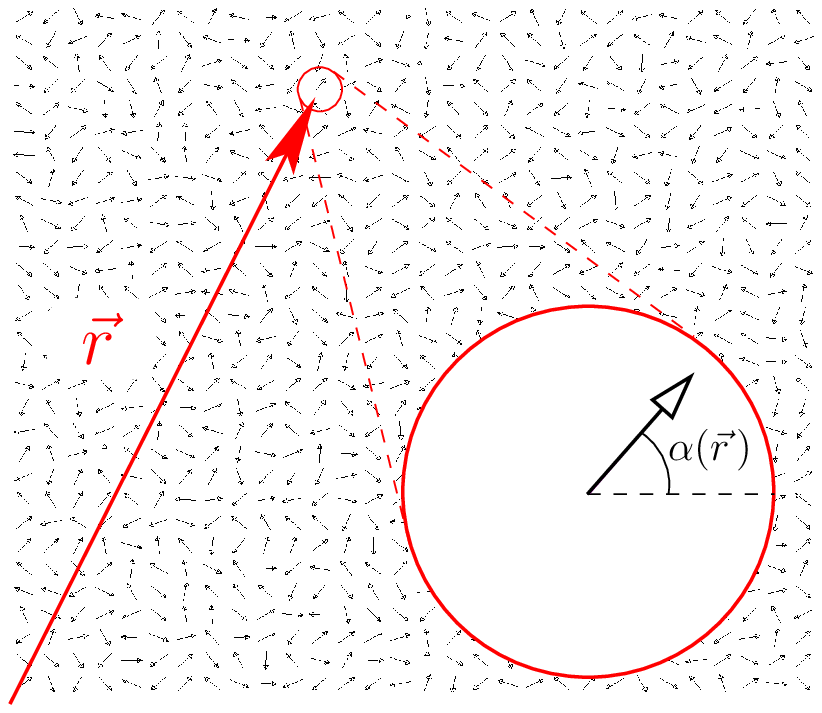}
\caption{\label{fig:modeleXY_def} The \textsc{xy}-model describes an interacting bidimensional vector field
of constant and uniform norm.
On a continuous space or on a lattice, the direction of the field at point~$\vec{r}$ is
given by one angle $\alpha(\vec{r})$.  
 }
\end{center}
\end{figure}
\subsection{The \textsc{xy}-model}
The latter picture is known as the
\textsc{xy}-model, which is also relevant for some classical liquid
crystals or for systems of classical spins
\citep[chap.~6]{Chaikin/Lubensky95a}.  At macroscopic scales, some
collective effects of such model are not very sensitive to the details
of the interaction nor to the geometry of the elementary cell in the
case of a lattice but depend crucially on the dimension~$d$ of the
space of positions (the number of components of~$\vec{r}$).

\begin{figure}[!ht]
\begin{center}
\parbox{17cm}{\includegraphics[width=5cm]{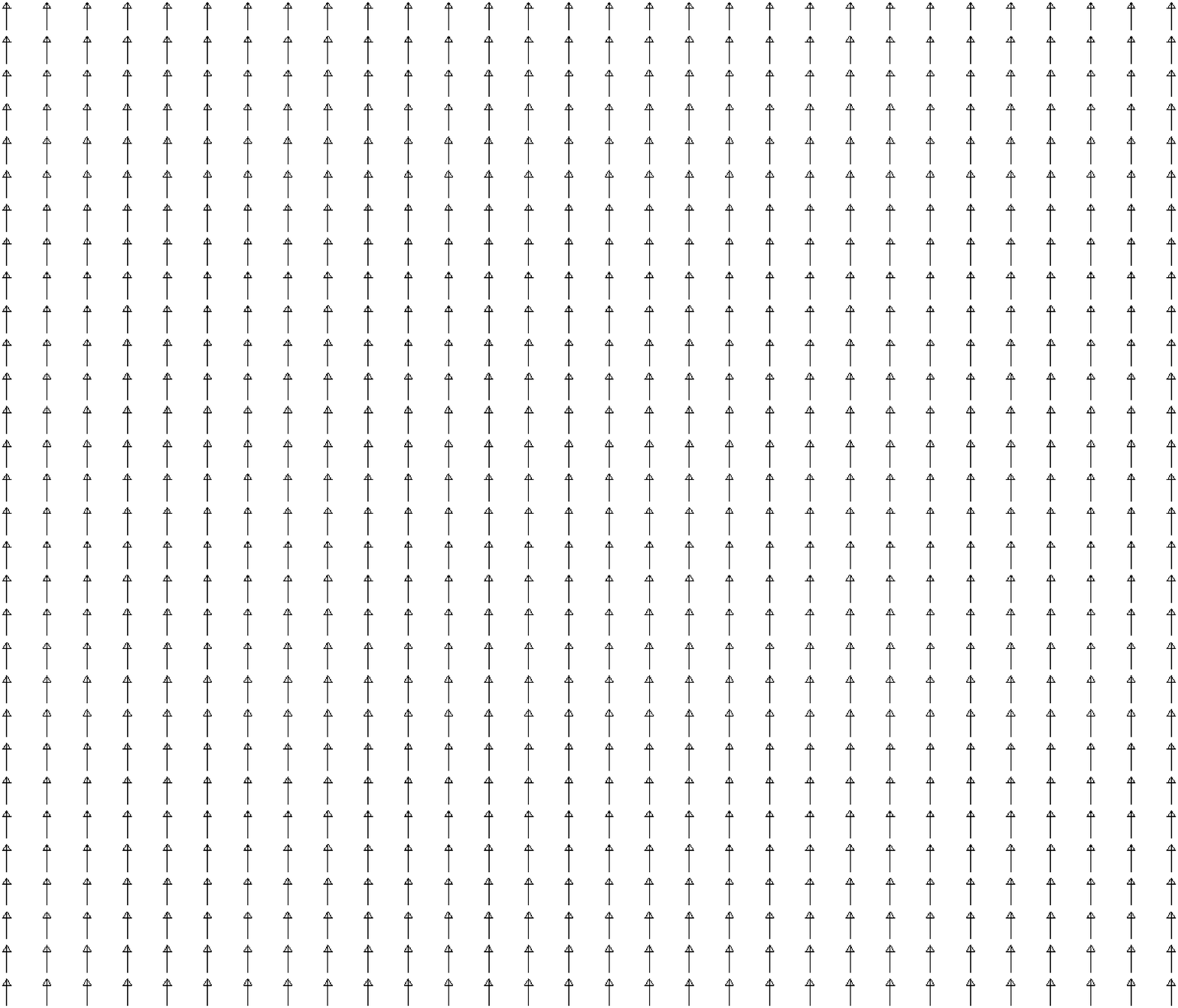}\ 
\includegraphics[width=5cm]{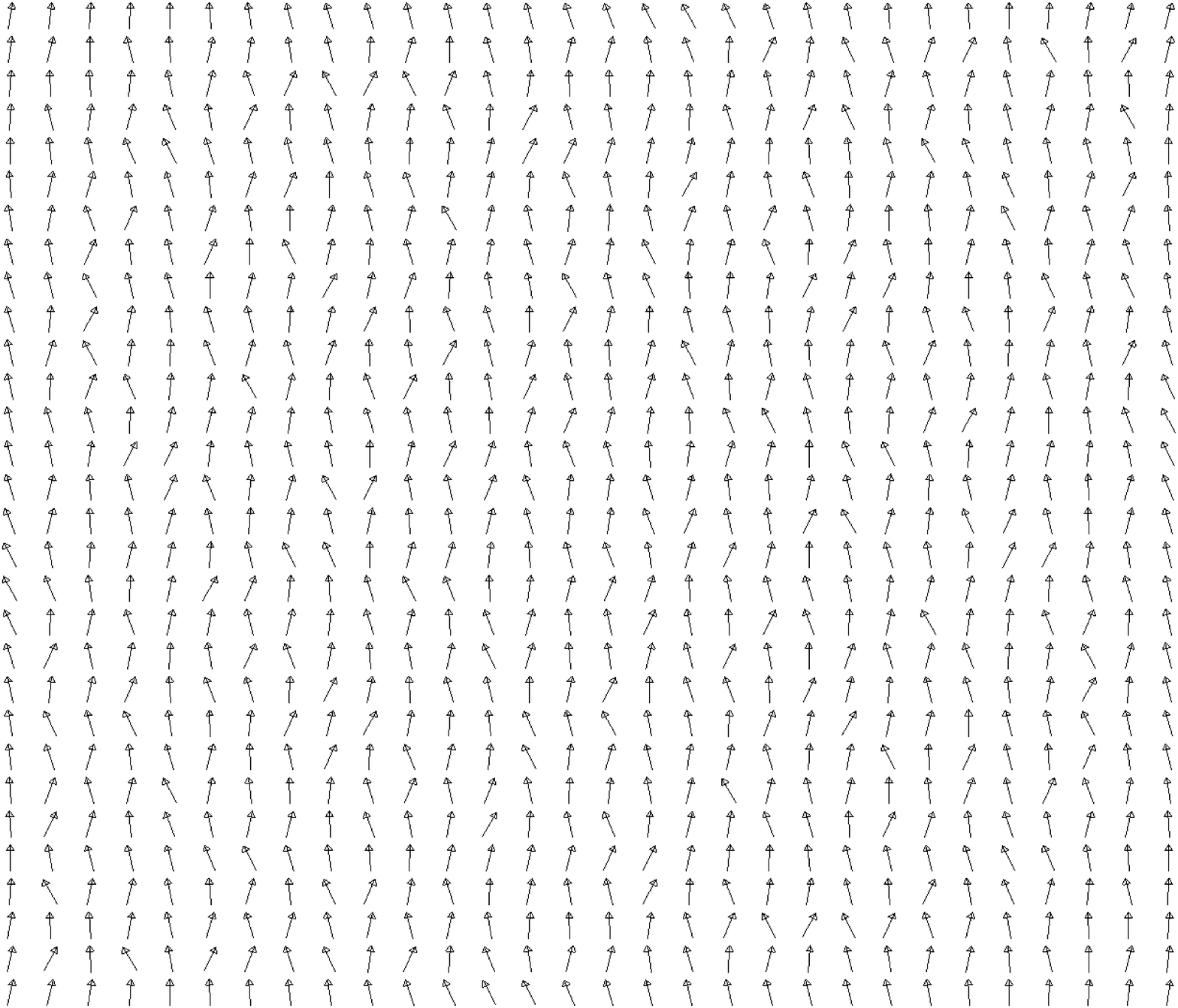}\ 
\includegraphics[width=5cm]{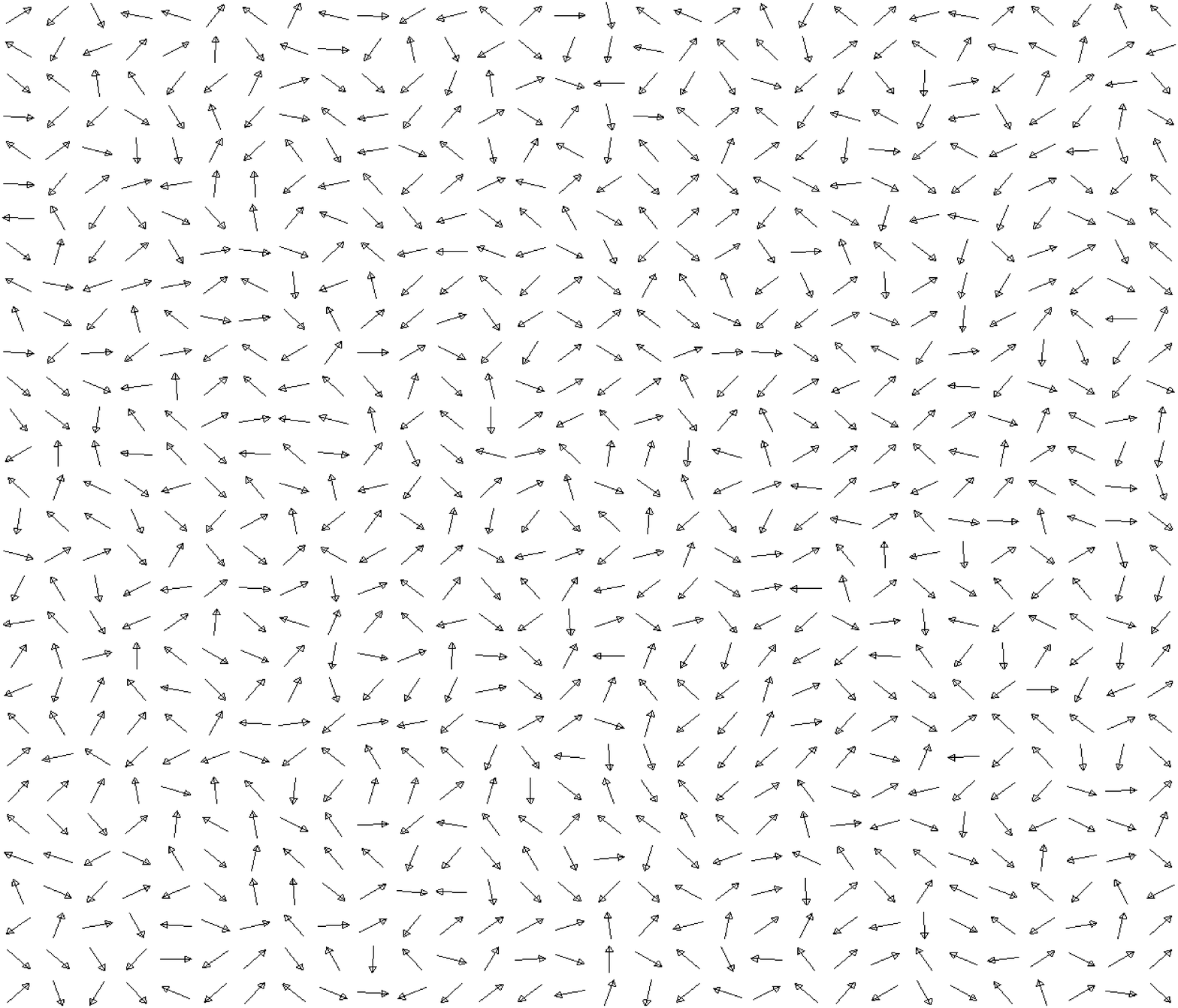}}
\caption{\label{fig:transitionordredesordre} In three dimensions the \textsc{xy}-model presents
order/disorder phase transition, very similar to the familiar solid/liquid phase transition.
Below a critical temperature~$T_{\mathrm{critical}}>0$ some order is maintained throughout the system
at macroscopic lengths (middle picture) with the perfect order obtained at~$T=0\mathrm{K}$ (left picture).
Above~$T_{\mathrm{critical}}$, the average orientation is zero and no more order at large scales can be identified (right picture).}
\end{center}
\end{figure}

Typically, the energy of the system increases when some differences in
the orientation~$\alpha$ appears; more precisely the energy density
contain a term proportional
to~$(\overrightarrow{\mathrm{grad}}\,\alpha)^2$. It is not affected by
a homogenous rotation of all the spins,
\begin{equation}\label{eq:uniformrotation}
  \alpha(\vec{r}) \mapsto \alpha(\vec{r})+\alpha_0\;,
\end{equation}
where the angle~$\alpha_0$ does not depend on~$\vec{r}$.  The absolute minimum
of the total energy is obtained when all the vectors are aligned,
which is the configuration at
temperature~$T=0\;\mathrm{K}$. When~$T>0$, the equilibrium corresponds
to more disordered configurations but,
for~$d=3$\footnote{Surprisingly, as far as the computations are
  concerned, the integer nature of~$d$ becomes secondary and one can
  formally consider~$d$ as continuous. The condition for an
  order/disorder phase transition at~$T_{\mathrm{critical}}>0$ to
  exist is~$d>2$. }, some non-zero average value of~$\alpha$ can be
maintained up to a critical temperature~$T_{\mathrm{critical}}$ beyond
which the average value of~$\alpha$ is zero
(fig.~\ref{fig:transitionordredesordre}).
At~$d=2$, on the contrary, the correlations
between fluctuations never decrease sufficiently rapidly at large distances 
and the average value of~$\alpha$ is zero as soon as~$T>0$. However one can still identify, 
at some  finite temperature~$T_{\mathrm{critical}}>0$, 
a qualitative change of behaviour in the correlation lengths, from a power-law decay at large distances to an exponential decay and this phase transition has observable repercussions, notably in
superfluids helium films~\citep{Bishop/Reppy78a}. The theoretical description
of what appeared to be a new kind of phase transition, now known as topological phase transitions,
  was proposed by \citet{Kosterlitz/Thouless72a}
who showed that vortices were a cornerstone of the scheme. 

\begin{figure}[!hb]
\begin{center}
\includegraphics[width=11cm]{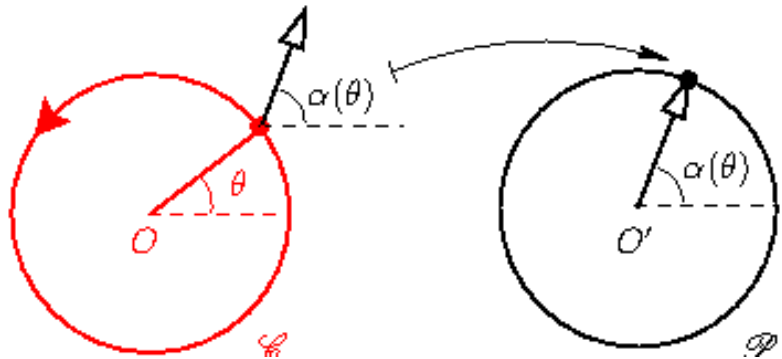}
\caption{\label{fig:modeleXY_index} For the \textsc{xy}-model, in~$d=2$,  to each point on a
loop~$\mathscr{C}$ 
enclosing any given point~$0$ (in red, on the left) is associated the direction of the 
order parameter on the circle~$\mathscr{P}$ (in black on the right).   }
\end{center}
\end{figure}

\begin{figure}[!hb]
\begin{center}
\includegraphics[width=7cm]{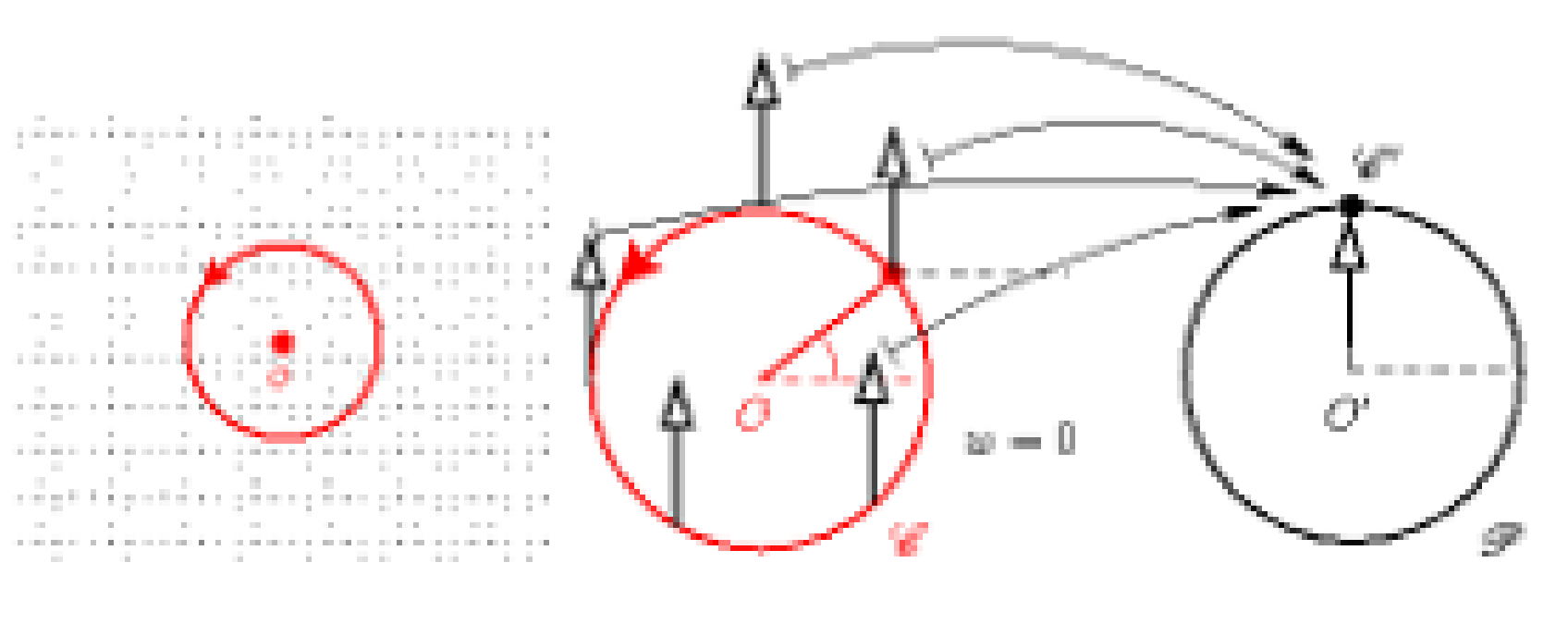}
\includegraphics[width=7cm]{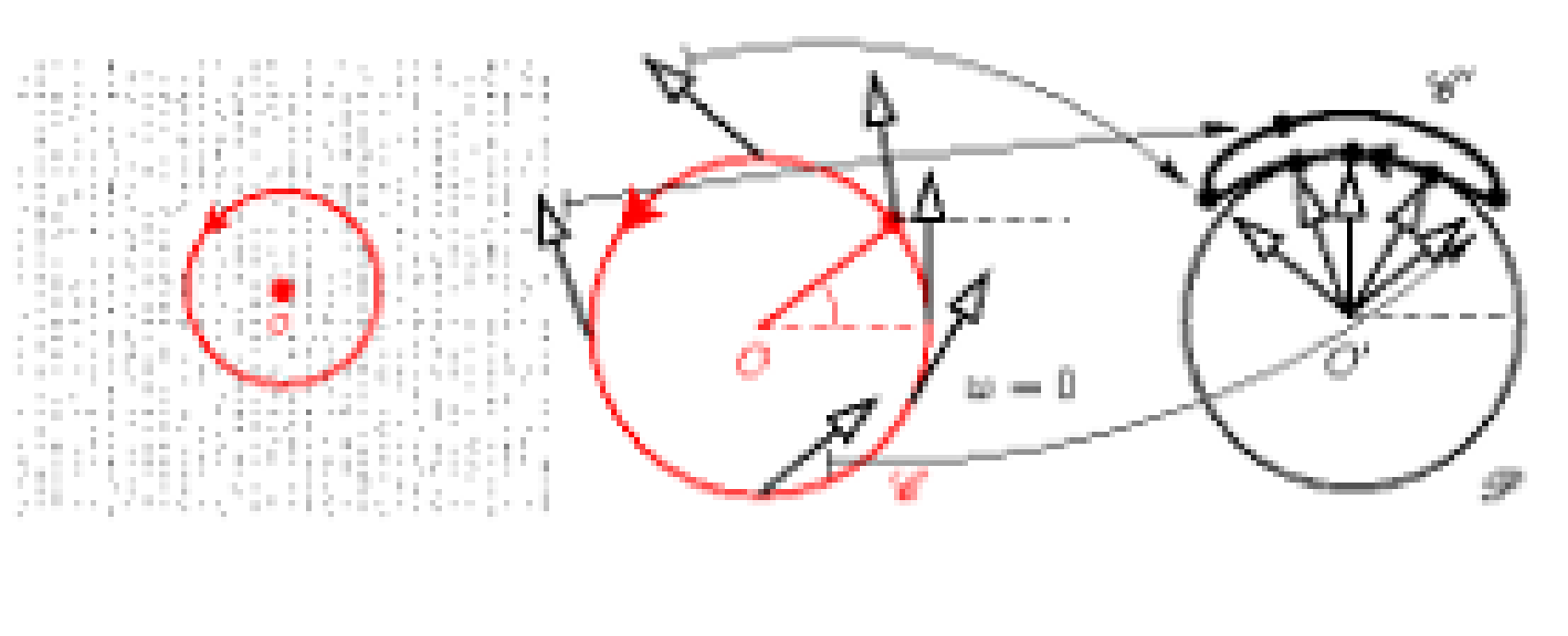}
\includegraphics[width=7cm]{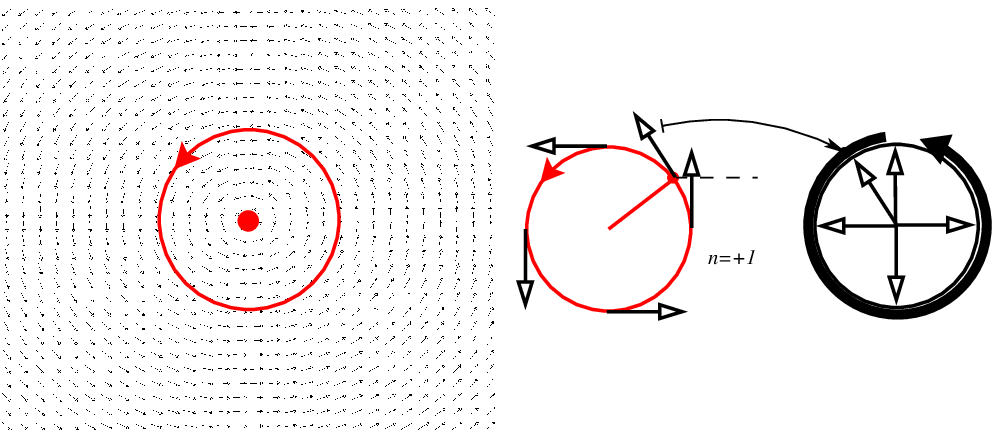}
\includegraphics[width=7cm]{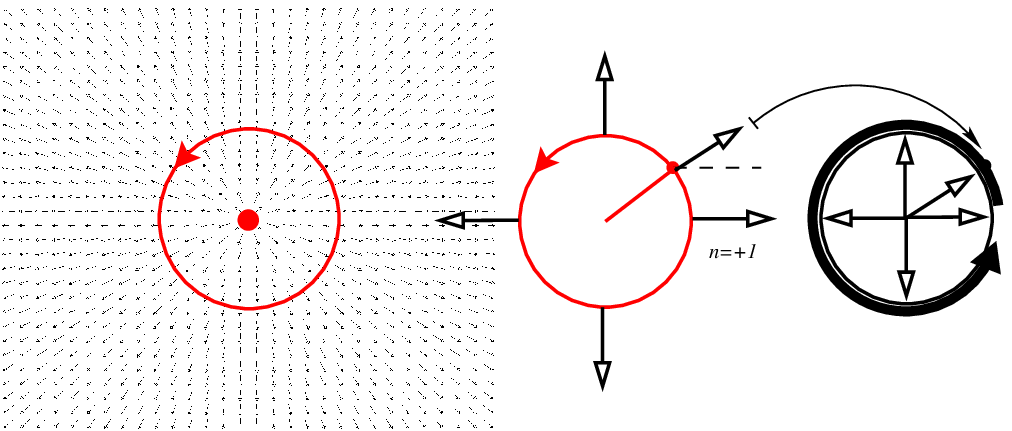}
\includegraphics[width=7cm]{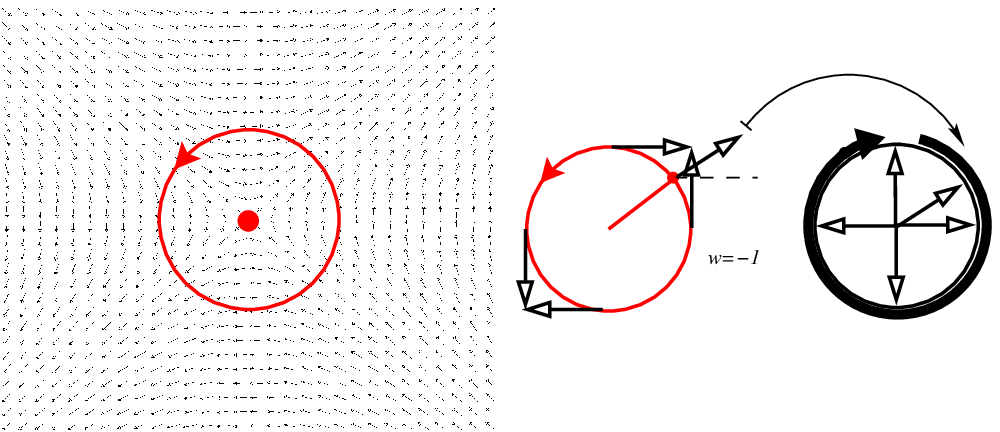}
\caption{\label{fig:xy_index_01m1} In the \textsc{xy}-model, the topological index~\eqref{eq:winding}
of a point~$O$ is the winding number of the curve~$\mathscr{C}'$ (thick black line) defined to be the image
of a closed loop~$\mathscr{C}$ (in red) in the circle~$\mathscr{P}$ (thin black circle) that indicate the direction~$\alpha$
of the order parameter. A smooth deformation 
deforms~$\mathscr{C}'$ but do not change~$w$ (we stay in the same homotopy class). The upper row provides two examples
having~$w=0$ (with, on the left, a uniform order parameter, $\mathscr{C}'$ is just a point).
The central row provides two elementary vortices ($w=1$) whose configurations differ from left to right by a 
rotation~\eqref{eq:uniformrotation} with~$\alpha_0=-\pi/2$. The lower row provides an example of configuration
having an elementary antivortex ($w=-1$).  }
\end{center}
\end{figure}

As soon as their first papers, Kosterlitz and Thouless, talked about
``topological order'' because they were perfectly aware that this type
of phase transition, unlike all the phase transitions known at the
time of their publication, relies on topology rather than on symmetry
(breaking). As we have seen above on eq.~\eqref{eq:winding}, each
vortex (now a topological defect of one dimension) is characterised by
an integer, called the topological index of the vortex which can be
reinterpreted using the concepts introduced by Poincar\'e in a series
of papers that can be considered as the foundations of topology as a
fully autonomous research discipline \cite[\S\;4]{Epple98a}.  Any
direction far away a topological defect of dimension~$f$ in a space of
dimension~$d$ is represented by an element of the rotation group
in~$n=d-f-1$ dimensions, in other words such a defect can completely
enclosed by a~$n$-dimensional sphere~$S_n$. In~$d=3$ dimensions a wall
(a surface of dimension~$f=2$) cannot be enclosed ($n=0$), a
vortex-line ($f=1$) can be enclosed by a circle ($n=1$), a point
($f=0$) can be enclosed by a $n=2$-sphere.  In~$d=2$ dimensions a wall
(a line of dimension~$f=1$) cannot be enclosed ($n=0$) and a point can
be enclosed by a circle ($n=1$). To each direction one can associate
the value of the order parameter and therefore to each defect one gets
a map from~$S_n$ to~$\mathscr{P}$ where~$\mathscr{P}$ denotes the
space to which the order parameter belongs. In the examples
above~$\mathscr{P}$ is just the set~$S_1$ of the angles~$\alpha$ but
much more different situations may be encountered.  For~$n=1$, any
loop~$\mathscr{C}$ around a given point maps on a closed
path~$\mathscr{C}'$ in~$\mathscr{P}=S_1$ and the topological index~$w$
of the point is just the winding number of~$\mathscr{C}'$
(figs.~\ref{fig:modeleXY_index} and \ref{fig:xy_index_01m1}).
\newpage


\newpage

\begin{figure}[!ht]
\begin{center}
\parbox{17cm}{\includegraphics[width=5cm]{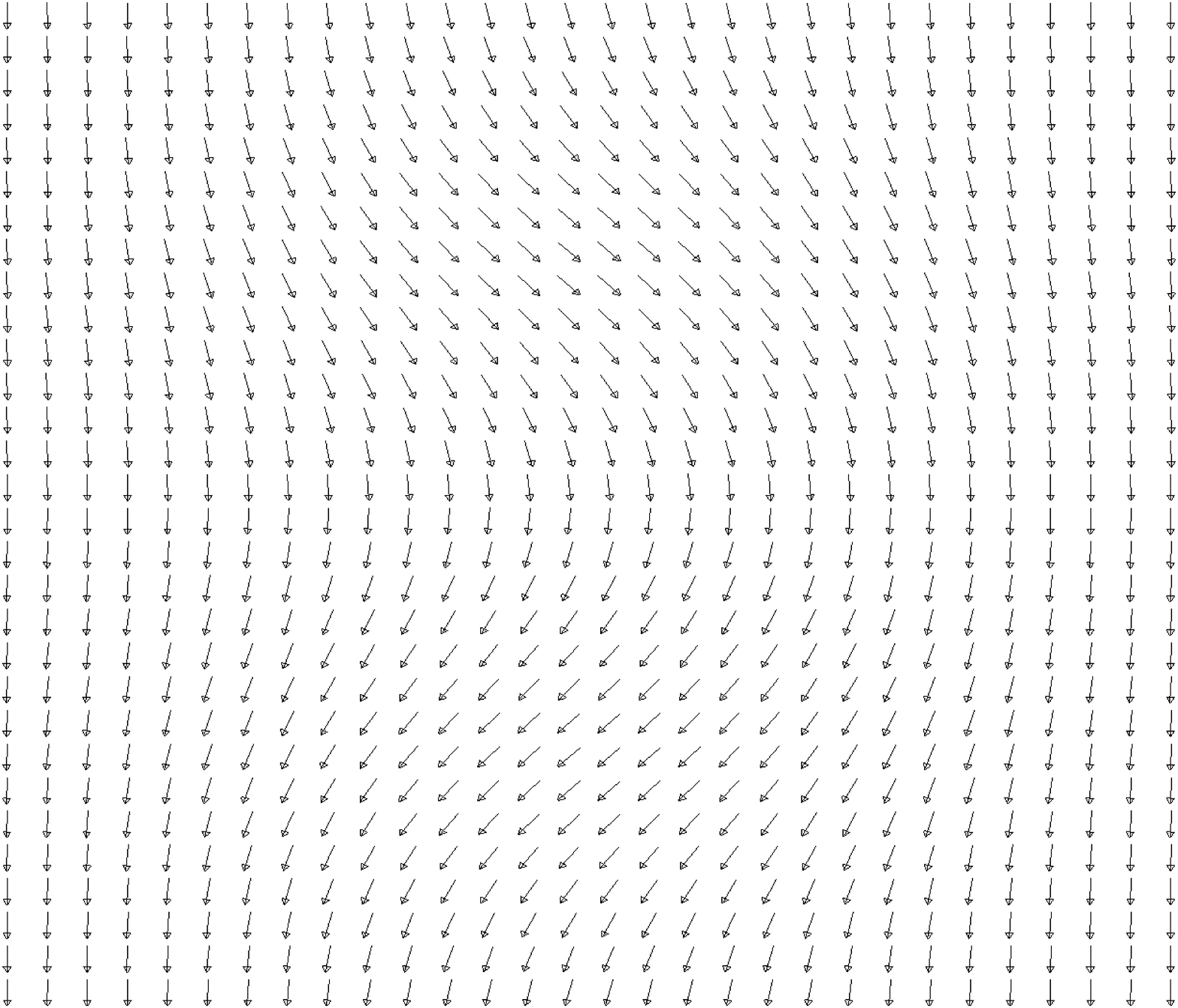}
\includegraphics[width=5cm]{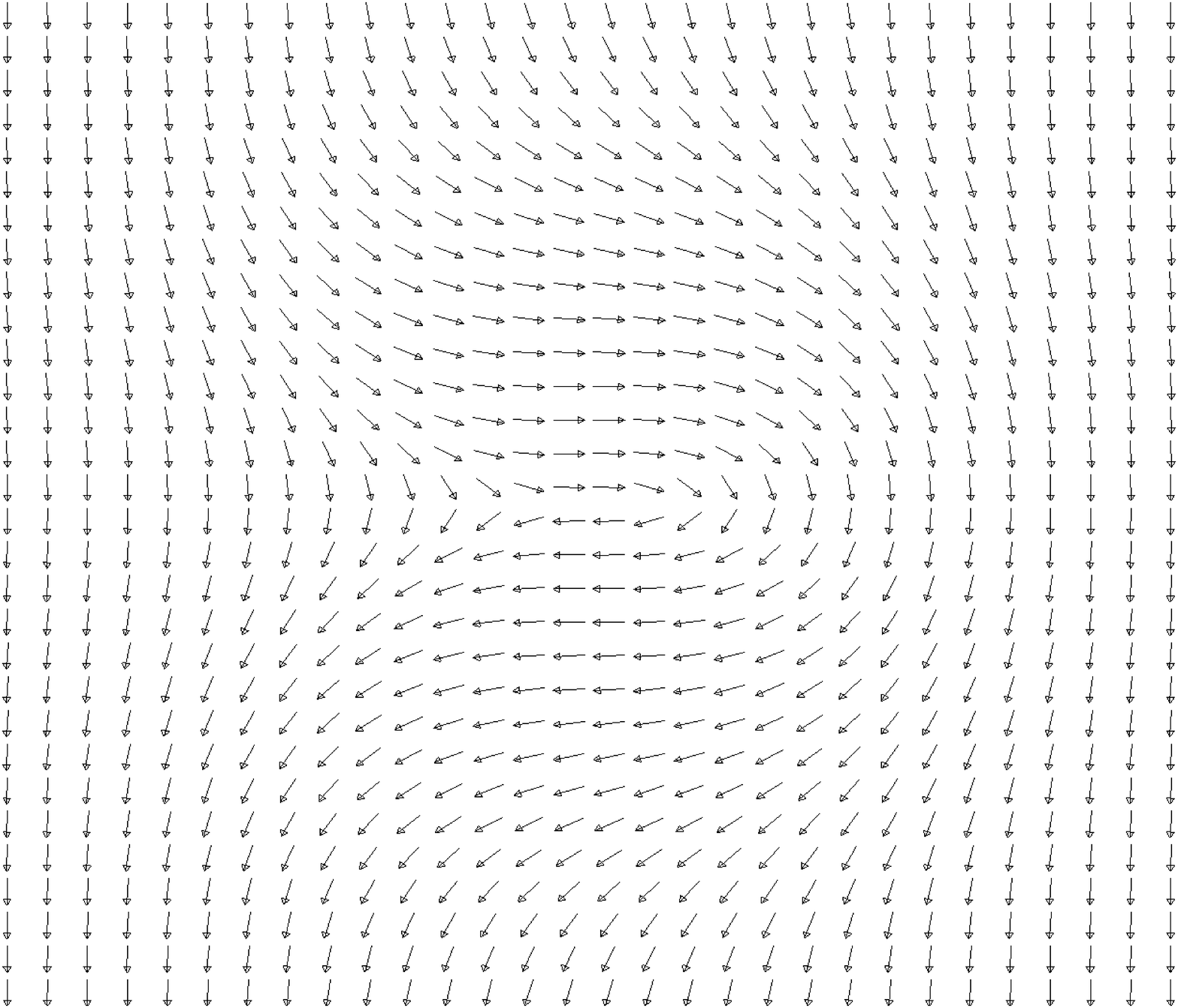}
\includegraphics[width=5cm]{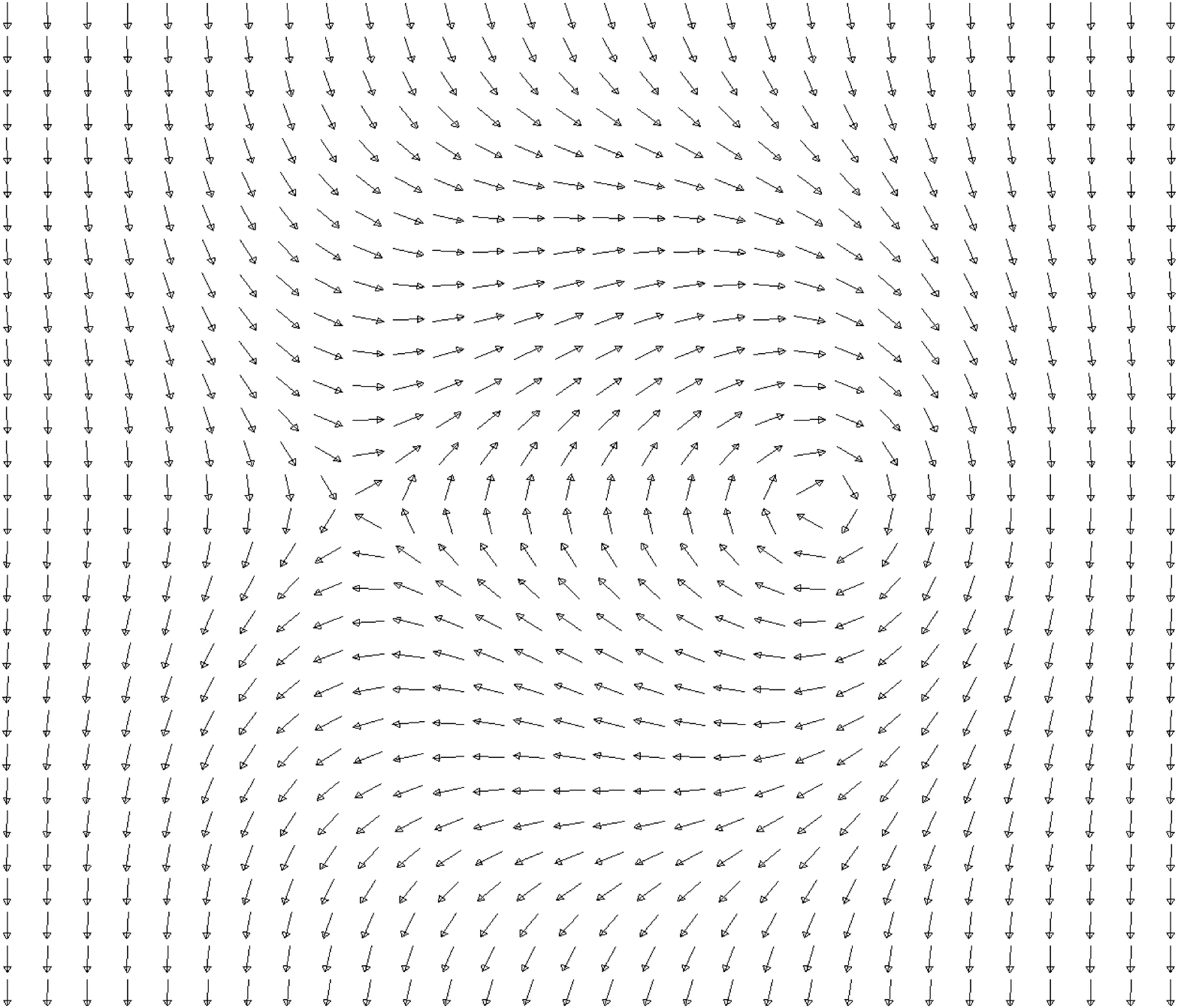}}
\caption{\label{fig:xy_index_pair} A smooth transformation that does not require a macroscopic amount of energy
can make a vortex/antivortex pair to spontaneously appear as a local fluctuation at non-zero temperature.
The genericity and the structural stability of this scenario can be understood when considering the 
appearance of a fold (fig~\ref{fig:pli}). }
\end{center}
\end{figure}
\begin{figure}[!ht]
\begin{center}
\includegraphics[width=\textwidth]{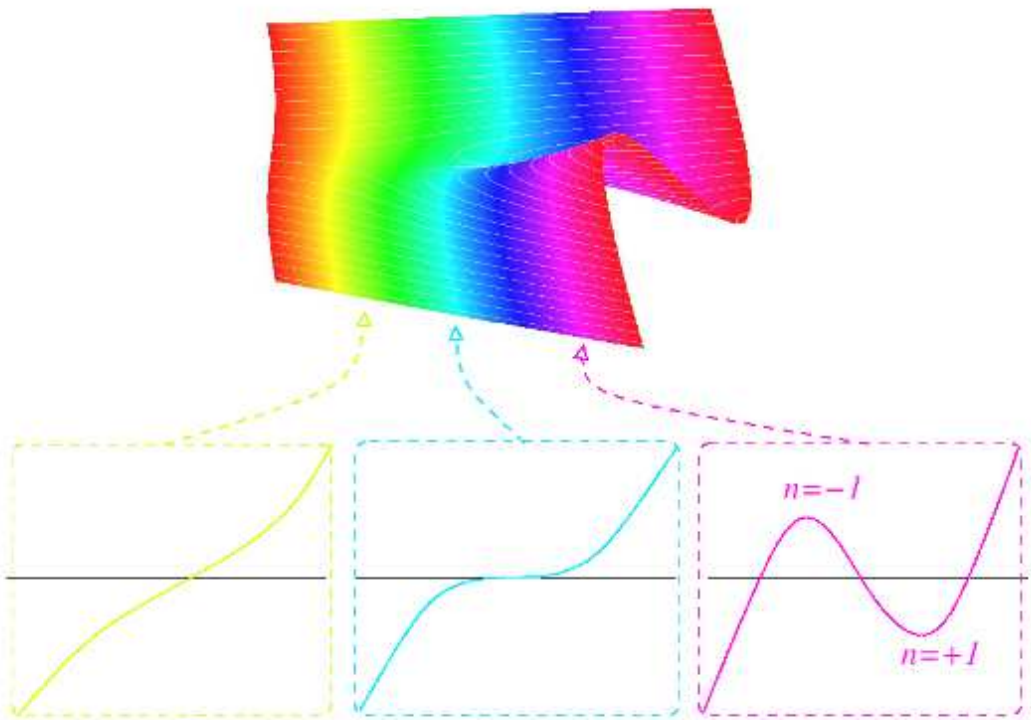}
\caption{\label{fig:pli} The fold catastrophe is the simplest of the bifurcation scenario. It involves
a one real parameter family of functions where two generic critical points (on the right), having opposite second derivatives,  merge into a degenerate critical point (central graph) and disappear (on the left). It also represents
how generically
a non transversal crossing between two tangent curves (in the center) is unfolded from one (on the left) to three (on the right)
transversal crossings.  
}
\end{center}
\end{figure}

\newpage
\ 

  More
generally, the topological invariants are given by the group~$\pi_n$
of~$\mathscr{P}$ (for~$n=0$ it provides the connectedness, for~$n=1$
it provides the first homotopy group that is the simple connectedness,
etc.).  A continuous transformation of the configuration cannot
modify~$w$ at any point and physically it would require a macroscopic
amount of energy to change~$w$. On the other hand, one configuration
having one defect can be deformed continuously at low cost of energy
into any other configuration having a defect with the same~$w$. In
particular, the transformation~\eqref{eq:uniformrotation} does not
cost any energy at all.

One cannot therefore expect to \emph{isolated} elementary vortex
($w=1$) or \emph{isolated} elementary antivortex ($w=-1$) to be
spontaneously created from a perfect ordered state. Nevertheless, a
pair of vortex-antivortex is affordable when~$T>0$
(fig.~\ref{fig:xy_index_pair}). The continuous creation (or
annihilation) of such a pair can be understood by considering the
appearance of a fold on a drapery (back to Leonardo again?).  One may
intuitively see that this is a generic process, stable with respect to
smooth transformations, that describes the creation or the
annihilation of a pair of maximal-minimal points on a smooth function
(fig.~\ref{fig:pli}) or, equivalently, the creation or annihilation
of intersection points when two curves that cross transversaly are
smoothly locally deformed\footnote{Topology is fully at work here and
  the study of the stability of the critical points of smooth mappings
  is the object of catastrophe theory whose greatest achievement is to
  have classified the generic possible scenarios; the simplest one
  being precisely the fold catastrophe, depicted in
  figure~\ref{fig:pli} \citep[for a general
    survey]{Poston/Stewart78a,Arnold84a}.}.  The topological phase
transition describes precisely how the creation of an increasing
number of vortex-antivortex pairs as the temperature increases
eventually lead from a topological order to a state where complete
disorder reigns.

\section{Concluding remark}

To come back to issues mentioned in the last paragraph of~\S\;1, in quantum
theory, the fundamental elementary particles stem from algebraic
symmetry considerations. However, we have some clues (topological
defects, solitons, instantons, monopoles, etc.)  that topology may
offer a complementary ground. The parallel between
creation/annihilation of particle-antiparticle pairs and
creation/annihilation of vortex-antivortex pairs may be more than a
simple analogy.  

 Thomson/Kelvin's intuition may take an unexpected
but relevant form, after all.

\bigskip

\textbf{Acknowledgement}: I am particularly grateful to Pascal Brioist  (Centre d'\'Etude Sup\'erieures de la Renaissance
de l'Universit\'e de Tours) for his expert advices on Leonardo studies, to Boris Behncke (\textsc{invg}-Osservatorio Etneo)) for letting me use
his photo of the vapour ring created by the Etna (see fig~\ref{fig:vortexanneau}) and to Michele Emmer who triggered the subject of this essay
for the conference \textit{Matematica e Cultura 2017}, Imagine Math~6, at Venice.

\nocite{James99a}


\end{document}